\documentclass[useAMS,usenatbib]{mnras}
\usepackage{graphicx}
\usepackage[T1]{fontenc}
\usepackage{color}
\usepackage{url}
\usepackage[]{amssymb}

\title[Peculiar Velocities and the Light Cone]{A Full Treatment
  of Peculiar Velocities on the Reionization Light Cone}
\author[E. Chapman et al.]
{Emma Chapman,$^1$\thanks{e.chapman@imperial.ac.uk}
Mario G. Santos,$^{2,3}$\\ 
$^1$Astrophysics Group, Imperial College London, Blackett Laboratory, Prince Consort Road, London, SW7 2AZ, United Kingdom\\
$^2$University of the Western Cape, Department of Physics and Astronomy, 7535 Bellville, Western Cape Province, Cape Town, South Africa\\
$^3$The South African Radio Astronomy Observatory (SARAO), 2 Fir Street, Black River Park, Observatory, 7925, South Africa}

\def\LaTeX{L\kern-.36em\raise.3ex\hbox{a}\kern-.15em
    T\kern-.1667em\lower.7ex\hbox{E}\kern-.125emX}

\begin{document}

\maketitle

\begin{abstract}
Accurate simulations of the 21-cm signal from the Epoch of Reionization require the generation of maps at specific frequencies given the values of astrophysical and cosmological parameters. The peculiar velocities of the hydrogen atoms producing the 21-cm radiation result in a shift in the observed frequency of the 21-cm radiation and changes the amplitude of the signal itself. 
This is not an effect we can remove but instead needs to be accurately modelled to ensure we infer the correct physical parameters from an observation. 
We describe the full calculation of the distortion of the 21-cm signal, and propose a new code that integrates the 21-cm intensity along the line of sight for each individual light cone pixel to fully capture the intensity contributions from multiple redshifts. This algorithm naturally deals with the typical divergences found in standard approaches, allowing for large optical depths and 21-cm absorption events at high redshifts. 
We find the new method results in up to a 60\% decrease in power on the largest scales on the sky, and an increase of over 80\% on the smallest scales on the sky. We find that the new implementation of the light cone results in a longer tail of bright temperatures in the brightness temperature distribution, as a result of the successful circumventing of a previous cap that had to be implemented to avoid a divergence in the brightness temperature. We conclude that this full treatment of the evolution of the light cone pixel can be an important effect.

\end{abstract}


\section{Introduction}

380,000 years after the Big Bang electrons and protons in our
Universe combined into neutral hydrogen and the Universe became
transparent to photons. As the Universe continued to expand and cool it
settled into its `Dark Ages', as the underlying dark matter structure
formed and the neutral hydrogen was gravitationally attracted to the
densest clumps of dark matter. Eventually, the density of this hydrogen was
such that fusion could begin and the first stars were
born. The Epoch of Reionization is the time following the Dark Ages
when the first ionizing sources in our Universe, whether mini-quasars,
stars or galaxies, began to emit radiation and ionize the
surrounding hydrogen. Eventually these patches of ionized hydrogen
grew such that they overlapped and the majority of the Universe was
ionized, by around one billion years after the Big Bang. 

Up until recently, the study of this field remained mostly theoretical,
apart from a few indirect constraints.  The Thomson optical depth as
measured by Planck can give us an integral constraint relating to the
transparency of the Universe to photons. The latest constraint is
$\tau=0.054 \pm 0.007$ \citep{Aghanim2018}, which is rather lower than
previous estimates from WMAP, relieving the tension from previous higher
values and setting an upper constraint on the extent of reionization to $\Delta z < 2.8$. Quasars are another key probe of reionization. Quasar spectra are sensitive to intervening neutral hydrogen with the Lyman-alpha (Ly$\alpha$) resonance in particular causing absorption lines in the broadband spectra of the quasar. These lines combine over redshift to form the Ly-$\alpha$ forest or, at higher redshifts when the forest saturates, a Gunn-Peterson trough, giving us an indication of the neutral fraction of the Universe \citep{Banados2017,Mortlock2011,fan06b}. Quasar absorption lines can be used to study the global ionizing emissivity budget, metal buildup and timing of reionization together implying that reionization comes to an end around $5.7 \leq z \leq 7.7$ \citep{Becker2015}. More specifically, quasars observed at very high redshifts can provide their own bounds for the neutral fraction at those redshifts, albeit along that line of sight only (e.g. \citet{Banados2017}, \citet{Mortlock2011}).

The 21-cm spin-flip transition of neutral hydrogen, either
in the form of an absorption or an emission relative
to Cosmic Microwave Background (CMB) blackbody
spectrum, holds the promise of revealing the detailed
astrophysical processes occurring during the Epoch of
Reionization (EoR) and the Cosmic Dawn (CD). 
For a review see e.g. \citet{furlanetto06a,pritchard10a, 2013ExA....36..235M}.
Several low frequency radio telescopes plan to observe this signal in order to detect the EoR (e.g. LOFAR\footnote{http://www.lofar.org/}, MWA\footnote{http://www.mwatelescope.org}, GMRT\footnote{http://gmrt.ncra.tifr.res.in}, PAPER\footnote{http://eor.berkeley.edu}, HERA\footnote{http://http://reionization.org} and the future and SKA1-LOW\footnote{http://skatelescope.org}). This neutral hydrogen signal extends back into the higher redshifts, when the first ionizing objects are still forming, in the Cosmic Dawn. Recently, a claim has been made of a possible detection of the absorption feature in the 21-cm global signal from the Cosmic Dawn by the EDGES telescope\footnote{https://www.haystack.mit.edu/ast/arrays/Edges/} \citep{Bowman2018}. This detection is still open to interpretation and an independent confirmation is required to ensure the absorption signal is real and not, for example, a conspiracy between the frequency dependence of the instrument and foreground contamination. Nevertheless, the wealth of information that we soon expect to access from all these experiments makes it the all more important to make sure we have an accurate calculation of the observed signal.

The 21-cm signal depends on several ingredients that can be calculated through simulations, in particular the fraction of neutral hydrogen and the spin temperature (more relevant at higher redshifts). 
Several codes have been developed for this, either full radiative transfer (e.g. \citealt{Gnedin2000,Finlator2009, Ciardi2003, Mellema2006, Semelin2007,Trac2007, Iliev2014}) or fast numerical simulations (e.g. \citealt{mesinger11,Mesinger2007,Zahn2007,Choudhury2009,2008ApJ...689....1S, santos2010}). However, turning these values into the observed 21-cm signal requires in principle a full radiative transfer calculation of the background radiation along the line of sight which depends on subtle and interconnecting effects. Firstly,
the peculiar velocities of the hydrogen atoms result not
only in the red-/blue-shifting of the 21-cm radiation but a boost/suppression of
the signal by a significant magnitude \citep{Majumdar2013,jensen13,mao12}. 
The resulting anisotropy in the power spectrum has the potential to reveal
rich cosmological and astrophysical
information (e.g. \citealt{barkana05}, \citealt{mao12}, \citealt{santos06, 2011A&A...527A..93S},
\citealt*{majumdar12}, \citealt{shapiro13}) and so needs to be
carefully understood.
Another observational effect is the cosmological
redshift of the 21-cm photons due to the expansion of the Universe,
$\lambda_{obs} = \lambda_{21} (1+z)$, which results in a set of observations
in frequency corresponding to not only different positions in space but also to different times in the evolution of the Universe. This is the so called light
cone effect. In order to construct a light cone one must take only the information from photons which have traveled the same light travel time. Analyzing a given observed volume without taking this evolution over time into account will produce biased results.

Both the light cone effect \citep{laplante14,Datta2012,Datta2014,Zawada2014,Barkana2006} and
the effect of the peculiar velocities on the 21-cm signal \citep{Bharadwaj2004,Majumdar2015,Ghara2015,mao12,Majumdar2013,jensen13} have been well studied. However, the meeting of the two is non standard. The usual approach is for brightness temperature cubes to be created at a given redshift with the peculiar velocity contribution included in the calculation and then, in a separate step, a light cone is assembled. This approach leads to problems when the peculiar velocity gradient is large, even leading to infinities that have to be dealt with in an ad hoc way. Moreover it doesn't take into account the fact that there could be contributions from multiple real space simulation box slices to the same map since the light cone map and redshift space distortion effects are coupled. \citet{mao12} describes an algorithm to deal with the peculiar velocity effects and the above "infinities" but although the coupling to the light cone is mentioned in Sec. 6.2.2, it wasn't implemented. One recent exception is \citet{Mondal2017} who divided the simulations into thin shells, re-assigning 21-cm events to an observed frequency including the effects of peculiar velocities. However, their approach does not consider that a 21-cm event may only partially contribute to a frequency map, having redshifted out of the observed frequency range as the Universe expands, and they also only consider the signal in emission at low redshifts, when the spin temperature is much larger than the CMB.

In this paper, we revisit the effect of the peculiar velocities on the light cone and present a new algorithm that naturally deals with the possible large gradients in the peculiar velocities, avoiding the issue of infinities in the brightness temperature altogether. The code allows for multiple scattering events along the line of sight (e.g. large optical depths) and evolution of the physical quantities along the line width when such width translates to a large time interval in the evolution of the Universe. Moreover, it includes the calculation of these effects during absorption, at high $z$, and is appropriate for fast semi-numerical simulations.

In Section \ref{sec:21-cm} we review the basic physics of the 21-cm radiation, including the standard 21-cm brightness temperature calculation, introduce the light cone effect and discuss
the common way of including this effect, uncoupled to peculiar velocities. In Section \ref{sec:21-cm_new} we rederive the brightness temperature equation considering how the 21-cm intensity adjusted for redshift space distortions is observed due to the light cone effect, thus taking into account the inherently coupled nature of the two effects. The new open source algorithm resulting from this new derivation is compared with other methods in the literature in Section \ref{comparison}. We summarise the semi-analytic
code which we are adapting in Section \ref{simfast21}. In Section \ref{results} we compare images and power spectra of the new algorithm with the traditional uncoupled treatment of the light cone and peculiar velocity effects.

\section{The 21-cm signal - standard case}
\label{sec:21-cm}

We start by re-deriving the fundamentals of the 21-cm signal for the
standard case. Although this has been discussed previously in the literature (see e.g. \citealt{furlanetto06a}), we believe it will be helpful to review the steps again in detail, in order to make it clear what approximations are usually made and what is new in the algorithm we are introducing.

The 21-cm signal corresponds to the intensity, $I_{21}(\nu_0)$, observed at frequency $\nu_0$, due to the combined effects of emission and absorption of radiation along the line of sight by neutral hydrogen from its 21-cm line. This line is particularly useful to probe the high redshift Universe, because of the abundance of neutral hydrogen as well as the weak energy associated with the line, which can be easily excited without the need for strong (and therefore rare) background sources. The observed intensity is usually considered against the background radiation, $I_{\rm bg}(\nu_0)$ (the radiation one would observe if there was no intervening HI), e.g. $I_{21}(\nu_0)-I_{\rm bg}(\nu_0)$, so that it can be positive or negative depending on the amount of background radiation absorbed versus the radiation emitted due to the line transition. Although some lines of sight might have a strong background source, this should be very rare on cosmological scales and the majority of the signal will have the cosmic microwave background as the background. It is common to define the brightness temperature:
\begin{equation}
\delta T_b(\nu_0)\equiv I_{21}(\nu_0)\frac{\lambda_0^2}{2 k_B} - T_{\rm CMB}(z=0)
\label{Tb}
\end{equation}
where $\lambda_0$ is the wavelength of the observed radiation, $k_B$ the Boltzmann constant and $T_{\rm CMB}$ is the CMB temperature today (2.725 K). Note that the calculation neglects angular fluctuations in the CMB temperature itself as the fluctuations are smaller than the brightness temperature fluctuations. From now on we will concentrate on the observed intensity $I_{21}(\nu_0)$ noting that equation \ref{Tb} should be considered when referring to the 21-cm signal brightness temperature.

\subsection{Changes in intensity}

The basic quantity we need to calculate is the variation in intensity $dI$ as the "light ray" crosses a patch of neutral hydrogen (note that the basic radiative transfer relations we use can be found in standard textbooks, for example, \citealt{mihalas78}). This change can be due to three effects: i) absorption/scattering of photons in the 21-cm resonance along the patch; ii) emission of 21-cm radiation from the patch and iii) Doppler effects due to the expansion of the Universe and peculiar velocities. Throughout the paper, we will call a 21-cm emission/absorption along the path of the light ray towards the observer, a "21-cm event". Ignoring for the moment the last contribution, the change in intensity at a frequency ($\nu$) across an infinitesimal positive line element $ds$, can be written as:
\begin{equation}
\label{radeqn}
dI(\nu) = -k_{21}(\nu) I(\nu) ds + j_{21} (\nu) ds,
\end{equation} 
where $j_{21}$ is the emissivity coefficient and $k_{21}$ is the absorption coefficient. The first term on the right accounts for absorption and the last term for emission ($\nu$ is the frequency during the event, which
needs to be within the 21-cm line). We will always consider the changes along a given line of sight as the radiation propagates towards the observer. Therefore, $s$ will be the line of sight "proper" length, but starting at some far away distance and increasing towards the observer (e.g. positive $ds$ as it goes down the line of sight towards the observer). This is opposite from the standard definition of line of sight distance.

Following \citealt{furlanetto06a}, the emissivity $j_{21}$ and absorption coefficient $k_{21}$, can be written to a good approximation as:
\begin{equation}
j_{21} (\nu) \approx [1.6 \times 10^{-40} \; \mathrm{J\; s}^{-1} \;\mathrm{sr}^{-1}] n_{\rm HI} \phi_{21} (\nu )
\end{equation}
and
\begin{equation}
k_{21}(\nu) \approx (2.6 \times 10^{-19} \ {\rm K\ Hz\ m^2})  n_{\rm HI} \phi_{21}(\nu)/T_S,
\end{equation}
where $\phi_{21}(\nu)$ is the 21-cm line profile (with the integral over frequency normalized to one), $T_s$ the spin temperature and $n_{\rm HI}$ is the number density of HI atoms in total (in proper units unless otherwise stated). 
Note that both $k_{21}$ and $j_{21}$ are functions of frequency, time and spatial position. They are only non-zero when there is neutral hydrogen at the position we are considering and when the frequency is within the 21-cm line.

In an expanding Universe, the frequency of the light ray will redshift/blueshift due to the uniform expansion and the peculiar velocity. Again for an infinitesimal positive line element $ds$ (in proper units), going down the line of sight towards the observer, the frequency of the photon is adjusted by $d\nu$:
\begin{equation}
d\nu =- \nu \frac{ds}{c} \left[H(z) + \frac{dv}{ds}\right],
\label{dnu}
\end{equation}
where $H(z)$ is the Hubble parameter at the redshift considered and $\frac{dv}{ds}$ is the "proper" gradient of the peculiar velocity along the line of sight (both $dv$ and $ds$ are in proper units). Note that the gradient will hold the same value whether we take $ds$ increasing or decreasing away from the observer (since the axis will be reversed). For $|dv/ds| < H$, the term in brackets will be positive and $\nu$ will decrease when $s$ increases towards the observer, as expected (the usual redshift).

The intensity of the "light ray" will also change from an initial ($I_i$) to a final value ($I_f = I_i + dI$) according to this shift in frequency from $\nu_i$ to $\nu_f = \nu_i + d\nu$: $I_f(\nu_f) =  (\nu_f/\nu_i)^3 I_i(\nu_i)$. Using the equation above for an infinitesimal $ds$, we obtain:
\begin{equation}
dI(\nu) = - 3 I(\nu) \frac{ds}{c} \left[H + \frac{dv}{ds}\right].
\label{dI_freq}
\end{equation}
To first order in $ds$, we can take the frequency at the beginning or the end of the patch $ds$. Neglecting the peculiar velocity term, this translates into the standard evolution: $I\propto a^{-3}$ (where $a$ is the scale factor and we note that according to our current definition, we have $ds=c dt$). In principle, this shift in intensity along $ds$ should be added to equation \ref{radeqn}, though we will consider if this is necessary in the next subsection.

\subsection{Approximations}

Using equations \ref{radeqn}, \ref{dnu} and \ref{dI_freq} above, one can in principle integrate the effects on the light ray along each line of sight in order to calculate the observed 21-cm signal. There are however several approximations usually made in the literature which we can now incorporate:
\begin{itemize}
\item
Take the 21-cm line profile $\phi_{21}$ to be a top hat function centred at the $\nu_{21}$ frequency, e.g. $\phi_{21}(\nu) = \frac{1}{\Delta\nu_{21}}$ across the 21-cm line, where $\Delta\nu_{21}$ is the line width.
\item
Assume there is only one absorption or emission 21 cm event for a given observed frequency $\nu_0$. This would happen at redshift $z=\nu_{21}/\nu_0 -1$.
\item
Assume $|dv/ds|/H \ll 1$, $\Delta\nu_{21}/\nu_{21}\ll 1$ and any physical quantities (such as $j_{21}$ and $k_{21}$) are constant across the line width. 
\end{itemize}
From equation \ref{dnu}, the proper distance traveled by the light ray across the event can now be written as 
\begin{equation}
\label{DeltaS}
\Delta s=\frac{\Delta\nu_{21}}{\nu_{21}}\frac{cH^{-1}}{1+1/H \frac{dv}{ds}}.
\end{equation} 
where we are already taking the frequency along the event as the 21-cm line frequency. The natural width of the 21-cm line is quite small $\sim 10^{-16}$ Hz, but thermal broadening will increase this by a large factor, $\Delta\nu_{21}\sim \nu_{21}/c\sqrt{2 k_B T_K/m_H}$ (with $T_K$, the temperature of the gas and $m_H$ the hydrogen atom mass). During reionization, the gas will be heated to temperatures of $\sim 10^4$ K, which will generate line widths of up to $\sim 60$ kHz. Still, the quantity $\Delta\nu_{21}/\nu_{21}$ will be quite small ($< 10^{-4}$). The same cannot be said of $|dv/ds|/H$ as we will see later.

Going back to equation \ref{radeqn}, we can rewrite the change in intensity across the 21-cm event as
\begin{equation}
\label{radeqn2}
\Delta I(\nu) = j_{21} (\nu) \Delta s \left[1 - \frac{I}{S_{21}}\right],
\end{equation}
where we are now using $\Delta$ to emphasize that we are considering a non-infinitesimal patch since we are considering the event across the full line width as explained in the assumptions above. Here, $S_{21}\equiv\frac{j_{21}}{k_{21}}$ is the specific intensity of the 21-cm transition, which is related to the spin temperature through $T_S=S_{21}\lambda_{21}^2/(2 k_B)$. Putting the values together, we have
\begin{equation}
\label{radeqn3}
\Delta I  = [1.127\times 10^{-49} \mathrm{J}\, \mathrm{s}^{-1} \mathrm{Hz}^{-1} \mathrm{sr}^{-1}] \frac{c H^{-1} n_{\rm HI}}{1+1/H \frac{dv}{ds}} \left[1 - \frac{I}{S_{21}}\right],
\end{equation} 

One last approximation is related to the shift in intensity in an expanding Universe, according to equation \ref{dI_freq}. This is obviously the main effect when there is no 21-cm event but it needs to be checked if we can neglect it during the event itself. During the emission/absorption, the largest frequency shift will be the length of the 21-cm line. We can then write $\Delta I/I \sim 3 \Delta\nu_{21}/\nu_{21} \sim 10^{-4}$ (at most). For the situation where the initial radiation is the CMB, the contribution to emission from the 21-cm event will be $\Delta I/I \sim 5\times 10^{-2}$, which is much larger. If the $I/S_{21}$ term dominates, the contribution would be even larger. Therefore, it is safe to ignore the shift in intensity due to the shift in frequency across the event and we will do so throughout the rest of the paper. The bottom line for this is that, after we compute the emission/absorption across each event, we just need to rescale the intensity by the factor $(\nu_0/\nu_{21})^3$ to take into account the expansion effect (and take the initial frequency to be just the 21-cm one).

As described, the standard approach only assumes one event across the line of sight. Therefore the initial intensity is just the CMB one, e.g. $\Delta I = I_f - I_i = I_f - I_{\rm CMB}$. Using the Rayleigh Jeans approximation to transform the intensity into a temperature, the observed brightness temperature is finally:
\begin{eqnarray}
\label{21-cm_standard}
\delta T_b(\nu_0) &\approx& (1.82 \times 10^{-28} \ {\rm K\ m^2})\  \frac{n_{\rm HI}}{1+z}\times\\ \nonumber
&&\frac{cH^{-1}}{1+1/H \frac{dv}{ds}} 
\left[1 - \frac{T_{\rm CMB}}{T_S} \right],
\end{eqnarray}
where the relevant quantities are calculated at the redshift $z=\nu_{21}/\nu_0-1$. Using typical values, this will give an observed signal of about $\sim 14$ mK from emission at $z=10$ with the Universe 50\% ionized. 

The equation above represents the standard calculation used in practically all 21-cm papers and simulations. Using for instance the simulation box outputs at a given $z$ ($n_{\rm HI}$, $dv/ds$ and $T_S$) we can generate an observed $\delta T_b$ box for this same $z$. We can then perform any analysis on it, such as the power spectrum statistics. However, as discussed below, even in the framework of these approximations, such an approach is only valid for a small interval in redshift as otherwise there will be cosmic evolution (for instance, an 8 MHz interval corresponds to about $dz\sim 0.5$, where evolution can already be non-negligible).

\subsection{The standard light cone}
\label{sec:lc}

\begin{figure*}
\begin{minipage}{126mm}
\begin{center}
\includegraphics[width=120mm]{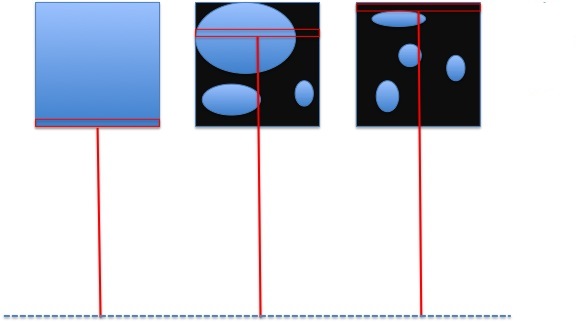}
\caption{A cartoon of a series of redshift simulation boxes with the $y$
axis the comoving line-of-sight. The boxes increase in redshift from left to
right, with the right-hand box representing $z_{max}$ and the left-hand box representing $z_{min}$. The red boxes represent the slices taken from each
box and the vertical solid lines represent the distance $x_{i}$ described
in the text.}
\label{lcex}
\end{center}
\end{minipage}
\end{figure*}

The so called light cone effect results from the fact that positions along the
line-of-sight are observed in the frequency direction. In this way, the observer cannot build a volume at a single moment in time. While simulations will output a 3D realization of the Universe at a chosen redshift,
observations will build up a volume where the light from positions farther away, corresponding to a lower frequency, will be from an earlier stage of reionization. In this way, statistical quantities constructed from the simulation outputs, such as the power spectrum, will only be an accurate representation of the observations for narrow slices where cosmic evolution can be ignored. A more accurate representation is to directly construct the light cone from simulations. The basic method employed to construct such a simulated observation cube from a series of simulation boxes is as follows (also refer to Fig. \ref{lcex}):

\begin{itemize}
\item Consider a series of simulation cubes representing the Universe
  between redshifts $z_{\rm min}$ and $z_{\rm max}$, all of comoving side length $L$
  and made up of $N$ cells along one side. We label the cells along the
  line-of-sight as $k_i$ where $0 \le i < N$ and $k_0$ is nearest to the observer. 
  The box comoving resolution is $\Delta r = L/N$. 
\item The simulation boxes all occupy the same region of comoving space in the Universe and we
  place the nearest edge (i.e. $k_{0}$) of all the boxes at a comoving
  distance $x_{min}$, which is the distance to the smallest output redshift, $z_{\rm min}$, with $x$ defined through:
\begin{equation}
x(z) = \frac{c}{H_0}\int^{z}_0 \frac{dz'}{\sqrt{\Omega_m(1+z')^3+\Omega_r(1+z')^2 + \Omega_{\Lambda}}}.
\end{equation}
In the absence of peculiar velocities, each frequency of observation, $\nu$,
corresponds to a simulation box output at a certain
redshift through the standard relation $1+z =
\frac{\nu_{21}}{\nu}$. If there is no simulation box at that precise
redshift, one can be obtained through interpolation.

\item From the box $z_{\rm min}$, we take the nearest slice along the
  line-of-sight, $k_0$, of the simulation
  box, and this becomes the slice at the highest frequency, $\nu_{\rm max}$, in our
  light cone.
\item For observed frequency $\nu_i$, the 21-cm signal corresponds to 
  redshift $z_i$:
$$1+z_i =
\frac{\nu_{21}}{\nu_i}$$ and at a comoving distance of $x_i$. We then
load box $z_i$ and take the slice at comoving distance $x_i$. This
will have a cell address of:
\begin{math}
k_i = \frac{(x_i - x_{\rm min})}{\Delta r} .
\end{math}
The slice located at line-of-sight address $k_i$ becomes the observed
frequency map at frequency $\nu_i$. 
\item In order to build up a light cone relating to observed
  frequencies between $\nu_{\rm min}$ and $\nu_{\rm max}$, a simulation box size $L$
  is required such that $L > x(z(\nu_{\rm min})) - x(z(\nu_{max}))$, though this
  can of course be satisfied by using boundary conditions.
\end{itemize}

We refer to the light cone
built up using this prescription from the real space brightness temperature boxes, as the standard light cone. 

\section{The 21-cm signal - full approach}
\label{sec:21-cm_new}

As we saw in the previous section, standard methods to calculate the 21-cm signal follow two separate processes:
\begin{itemize}
\item 
Calculate brightness temperature boxes using equation \ref{21-cm_standard}.
\item
Either ignore cosmic evolution and just proceed the analysis using the box at a given average $z$ or slice different boxes to construct a light cone as described in section \ref{sec:lc}.
\end{itemize}

This approach ignores that formally the brightness temperature is only well defined from the point of view of the observer, e.g. the signal does not exist in isolation at any redshift. Therefore, a proper algorithm should generate the line of sight cone and brightness temperature at the same time.
Moreover, as we can see in equation \ref{21-cm_standard}, the result is not well defined when $1/H \frac{dv}{ds} = -1$ which can occur in high density places. In fact, even for other values close to $-1$ the situation is ill defined as we will discuss later, so that a more in-depth algorithm should be used.

\subsection{Integration along a line of sight}
\label{sec:21-cmexp}

To calculate $I_{21}(\nu_0)$ from a given direction $\hat{n}$, we should consider the initial background radiation, $I_{\rm bg}(\nu)$ at some high redshift $z$ and then integrate the effects of absorption and emission along the line of sight $\hat{n}$ from redshift $z$ to $0$. The chosen starting redshift should be high enough to guarantee that there is no 21-cm event before it, so that every event is counted. The initial intensity will be at a comoving distance from the observer given by $x(z)$ as above.

As we integrate along the line of sight from $z$ to $0$ (or from $x(z)$ to $x=0$), we need to calculate the small changes in intensity $dI$ across the small steps $dx$. This is usually referenced instead in terms of the proper physical distance $ds=dx/(1+z)$. The change $dI$ will depend on the position $s$ (or redshift) and incident intensity as well as the frequency. The question is then how to generate the signal for a given angle, $\hat{n}$ and observed frequency, $\nu_0$. Our approach goes through the following steps:
\begin{itemize}
\item
For a given observed frequency, $\nu_0$, consider a high enough initial redshift ($1+z_i > \nu_{21}/\nu_0$) so that we are certain that there is no 21-cm event above that redshift contributing to this frequency (it is far away from the 21-cm line). The actual value was obtained by comparing to simulations and will be discussed later.
\item
The initial redshift will correspond to a comoving distance $x(z_i)$. We then go down along the line of sight towards the observer, looping through $x_{j+1}=x_j-dx$ for a fixed, positive $dx$. This corresponds to a proper interval $ds=dx/(1+z)$. Redshifts will change accordingly: $z_{j+1}=z_j-H\frac{ds}{c} (1+z_j)$.
\item
The initial starting frequency is given by $\nu_i = \nu_0 (1+z_i) [1+(v_i-v_0)/c]$, 
where $v_i$ is the peculiar velocity at the initial point and $v_0$ is the observer velocity which is just a dipole and can be ignored.
\item
The frequency is then updated through equation \ref{dnu}: $\nu_{j+1} = \nu_j \left[1 - \left(H(z) + dv/ds\right) ds/c\right]$.
\item
We then go through the line of sight until the frequency intersects with the 21-cm line and we have an event.
\item
The initial intensity is just: $I_i(\nu_i) = I_{CMB}(\nu_i)$. However, if there is no 21-cm event, the intensity will just change due to the normal expansion of the universe. Once there is a 21-cm event, the initial intensity (before any event) can be assumed to be the CMB value at the 21-cm frequency, $I_{CMB}(\nu_{21})$. It will then be updated according to equation \ref{radeqn} We can safely neglect the redshift effect on the intensity during the event as discussed in the previous section. This update is done for all the events along the line of sight, using the intensity value from the previous step.
\item
At low enough redshifts, $1+z_f < \nu_{21}/\nu_0$ (once we are sure there will be no more events) we stop the integration along the line of sight. The final intensity is then updated through $I_{21}(\nu_0) = (\nu_0/\nu_{21})^3 I_f(\nu_{21})$, where $I_f$ can be the result of several 21-cm events (integrated using equation \ref{radeqn}). Note that at any event, the frequency for $I$ will always be $\nu_{21}$ so that no frequency update is needed between events, except for the final update when moving to the observed frequency $\nu_0$.
\item
The final brightness temperature is then calculated using equation \ref{Tb}.
\end{itemize}

\begin{figure}
\includegraphics[width=80mm]{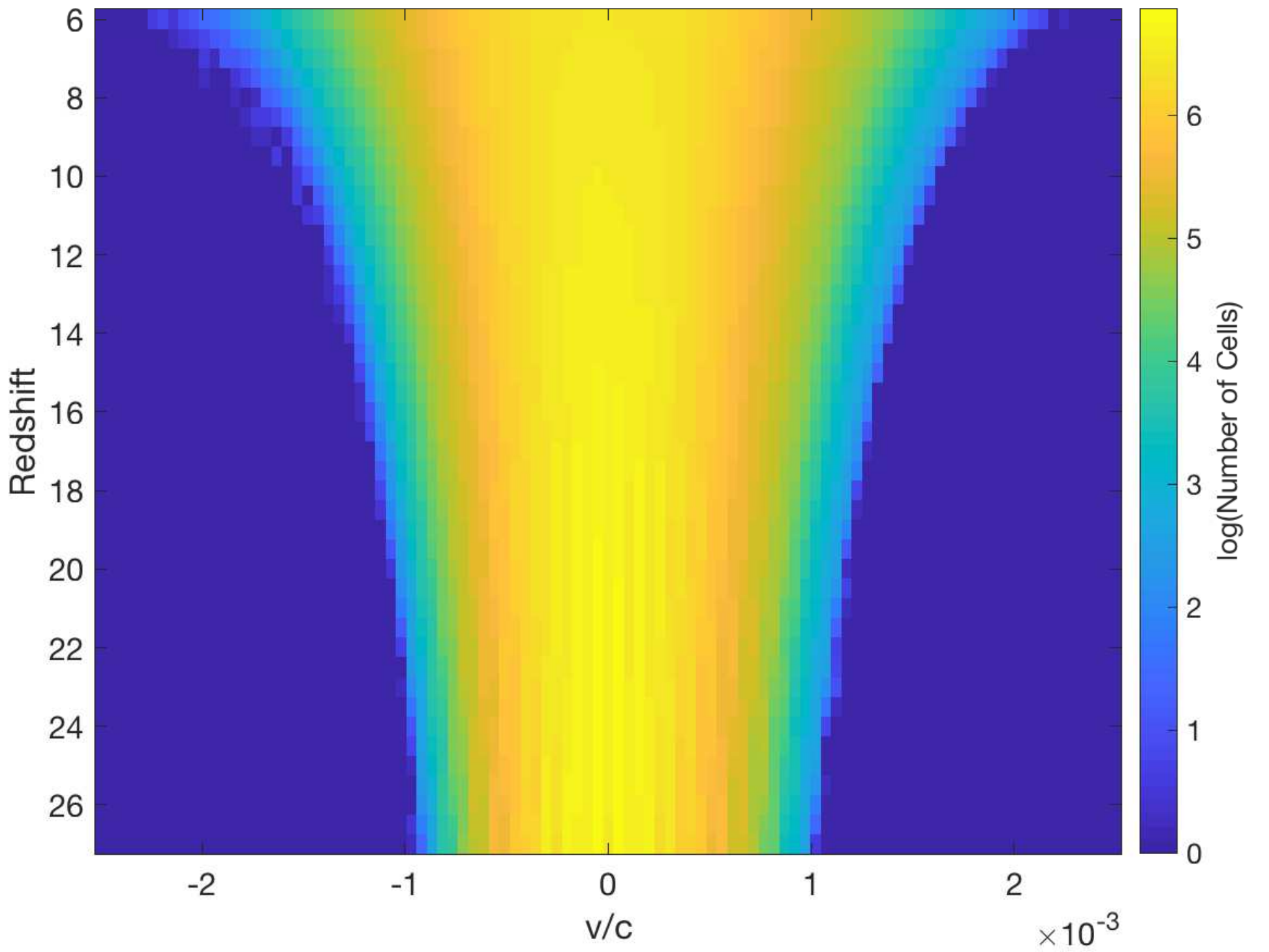}
\includegraphics[width=80mm]{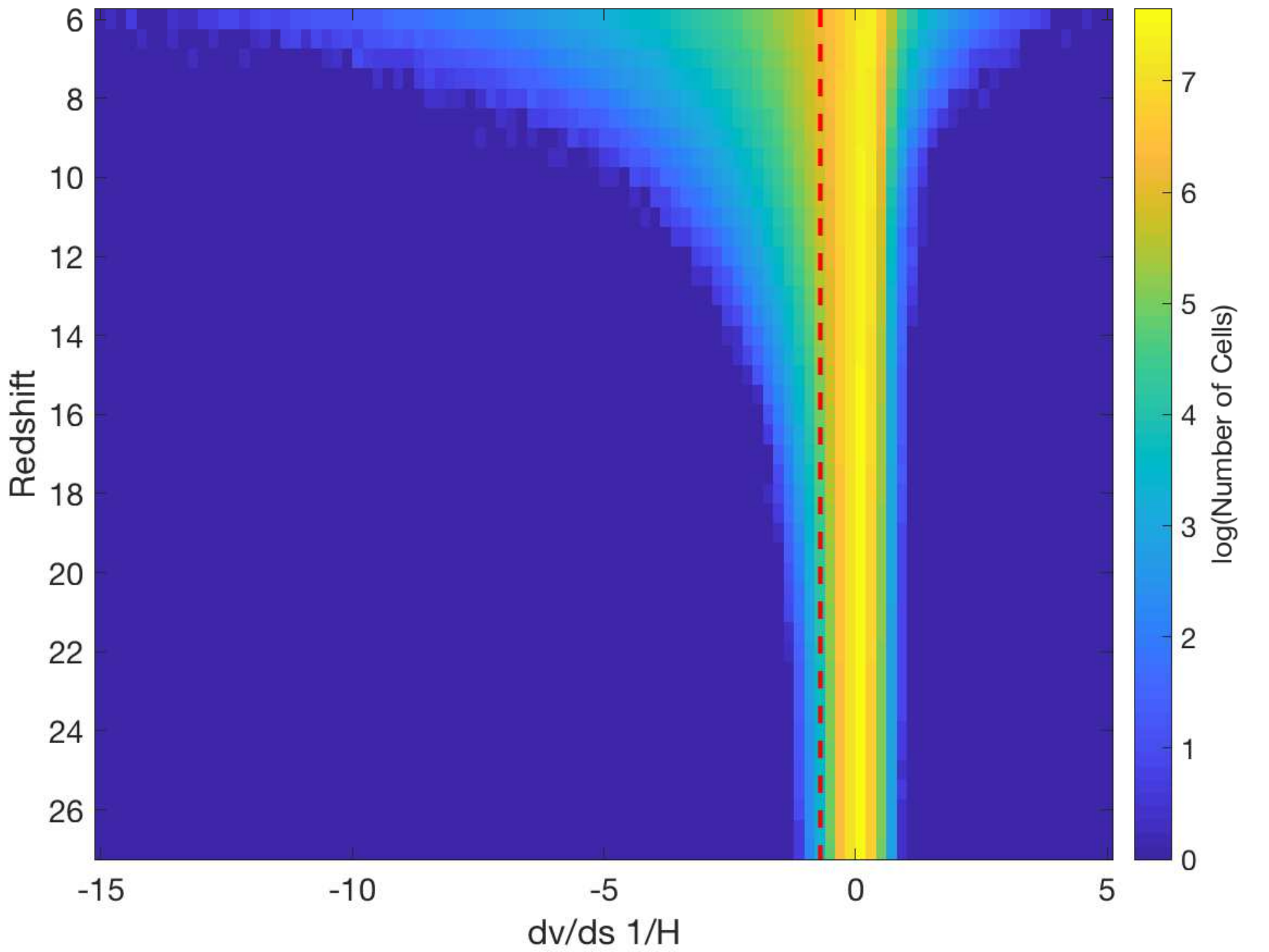}
\caption{Top: A heat map of the peculiar velocities, in the $z$ spatial
  direction, of all the simulation boxes
  at each redshift. The other two spatial directions are not shown as they are almost identical. Bottom: The same but for the gradient of the peculiar velocities, where the red dashed line represents the velocity gradient cut off required in \textsc{simfast21} to avoid divergence of the brightness temperature. The new light cone method can still incorporate all the information on the left of the red line. For both the velocity and gradient of velocities we see that the tails into larger values increase with decreasing redshift.}
\label{pvhists}
\end{figure}

For a given observing frequency, in principle we would need to do the calculation along the whole line of sight as peculiar velocities could move the frequency into the 21-cm line. This would be too slow and, as already indicated in the steps above, we only integrate between a given redshift interval. The target redshift is, as usual, $1+z=\nu_{21}/\nu_0$ and factoring in the line width and peculiar velocities, we only expect 21-cm events for an interval of:
\begin{equation}
\frac{\Delta z}{1+z}\approx \frac{\Delta\nu_{21}}{\nu_{21}}+2\frac{v}{c},
\end{equation}
where $v/c$ should correspond to the largest expected values of the peculiar velocity. This term dominates over $\Delta\nu_{21}/\nu_{21}$. In terms of the observed frequency resolution this would correspond to $\Delta\nu \sim \nu_{21}/(1+z)(2v/c)$. From the simulation, we get values at most of about $v/c\sim 1.5\times 10^{-3}$ (see Fig. \ref{pvhists}). At $z=8$ this would correspond to potentially filling maps in a $\Delta\nu \sim 0.5\ $MHz range. This will decrease with increasing redshift. To be conservative, we choose $\Delta\nu = 0.5\ $MHz throughout the simulation. We do not expect any 21-cm event outside this range. For a frequency map at $\nu_0 = 150$ MHz which would usually correspond to only a redshift of $z=8.47$, we will integrate events along the line of sight at points within the range $z=8.44 - 8.50$.

\begin{figure}
\includegraphics[width=80mm]{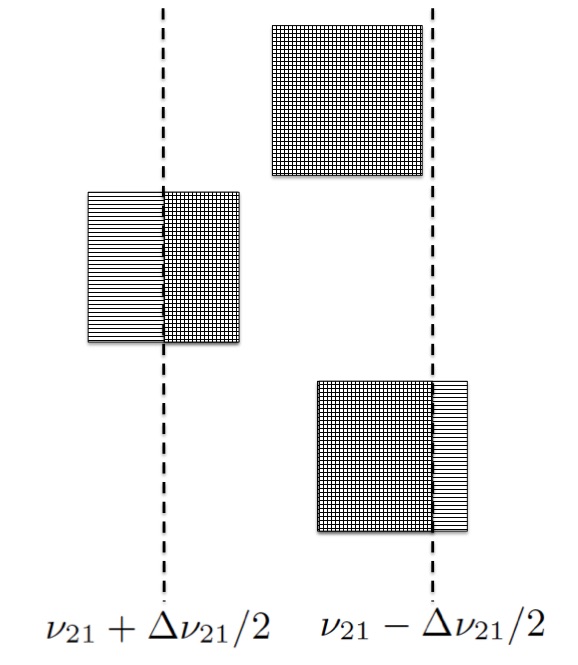}
\caption{A graphical representation of the possible intersections of $d\nu$ with the 21-cm line. The filled boxes represent the patch of $ds$ in frequency, i.e. $d\nu$, with the left side of the box being $\nu_{j}$ and the right hand side of the box being $\nu_{j+1}$. The cross-hatching represents where the 21-cm line (shown by the vertical dashed lines) intersects with the frequency patch - i.e. the portion of the patch ds over which a 21-cm event is observed. From top to bottom we have the cases where: the frequency patch is entirely within the 21-cm line; where the photon enters the line when within the patch; where the photon is already a 21-cm event upon entering the patch but redshifts out of it before getting to the end of the patch. For example for the third case we would have $\nu_{min} = \nu_{21} - \Delta \nu_{21}/2$ and $\nu_{max} = \nu_j$}
\label{im:conditions}
\end{figure}

One final issue is the actual calculation of the change in intensity using equation \ref{radeqn} since the step in frequency might be bigger or smaller than the 21-cm line width itself (since our step size is not infinitesimally small). A fixed comoving step size $dx$, can correspond to different frequency steps and $d\nu$ can be even zero if the velocity gradient $dv/ds = -H$. Again, the range of frequencies for which the 21-cm event will happen is set by the line profile, which we take as a top hat centered around $\nu_{21}$, with $\phi_{21}(\nu)=1/\Delta \nu_{21}$. Considering the intersection of the frequency patch with the line width, the maximum and minimum frequencies for the event  will depend on the frequency evolution. If $ \nu_{j+1} < \nu_j$ as it is usually the case, since we are going down in redshift, we have:
\begin{eqnarray}
\nu_{max} &=& \mathrm{min}\{\nu_j, \nu_{21}+\Delta\nu_{21}/2\} \\ \nonumber
\nu_{min} &=& \mathrm{max}\{\nu_{j+1}, \nu_{21}-\Delta\nu_{21}/2\} \nonumber,
\end{eqnarray}
If $ \nu_{j+1} > \nu_j$, we have instead:
\begin{eqnarray}
\nu_{max} &=& \mathrm{min}\{\nu_{j+1},\nu_{21}+\Delta\nu_{21}/2\} \\ \nonumber
\nu_{min} &=& \mathrm{max}\{\nu_j, \nu_{21}-\Delta\nu_{21}/2\} \nonumber.
\end{eqnarray}
These conditions are represented graphically for $\nu_{j+1}<\nu_j$ in Fig. \ref{im:conditions}.

The intensity during a 21-cm event is updated through:
\begin{equation}
\label{dI_main}
I_{j+1} = I_j+j_{21}(\nu_{21}) ds' \left[1-\frac{I_j}{S_{21}(\nu_{21})}\right],
\end{equation}
where the possibilities for $ds'$ are
\begin{enumerate}
\item 
if there is no event, e.g. neither $\nu_j$ and $\nu_{j+1}$ are within the line width, then $ds'=0$,
\item
if $d\nu = 0$, e.g. $\frac{dv}{ds} \sim -H$ (and $\nu_j$ is within the line), then $ds'=ds$,
\item
otherwise:
\begin{equation}
\label{ds_main}
\ \ \ \ \ \ \ \ ds'=\frac{\nu_{max}-\nu_{min}}{\nu_{21}}\frac{cH^{-1}}{|1+1/H \frac{dv}{ds}|}.
\end{equation}
\end{enumerate}

The situation above avoids the case when the patch goes to infinity as in the standard approach (equation \ref{radeqn3}). Note that $ds'$ can never be larger than the chosen $ds$. Moreover, by using a small $ds$ in the algorithm, we can implicitly avoid the situation where the simulation values can actually change across the patch. Finally note that when the line width is completely "inside" the patch, we have: $\nu_{max}-\nu_{min} = \Delta\nu_{21}$ and we recover equation \ref{DeltaS}.

\subsection{Required functions}

In order to implement equation \ref{dI_main} we need to have several ingredients: the hydrogen number density, $n_{\rm HI}$, the line profile, $\phi_{21}$, the specific intensity $S_{21}$, the peculiar velocity, $v$, as well as the peculiar velocity gradient $dv/ds$. As explained above, the line profile is assumed to be a top hat centred on the $\nu_{21}$ frequency and with width given by thermal broadening. Therefore, if we want to be more accurate in the use of the line width, we also need the temperature of the gas, $T_K$.

The specific intensity can be expressed in terms of the spin temperature, $S_{21}(z) = T_S(z) 2 k_{\rm B}/\lambda_{21}^2$ with
\begin{equation}
1-\frac{T_{\rm CMB}(z)}{T_{\rm S}(z)}=\frac{x_{\rm tot}}{1+x_{\rm tot}}\left(1-\frac{T_{\rm CMB}(z)}{T_{\rm K}(z)}\right),
\end{equation}
where $x_{\rm tot}=x_{\alpha}+x_{\rm c}$ is the sum of the radiative ($x_{\alpha}\geq 0$) and collisional ($x_{\rm c}\geq 0$) coupling parameters (for further details, see e.g. Section 2.4 in \citealt{santos2010}). At "lower" redshifts ($z\lesssim 15$), most implementations take $T_S(z)\approx T_K(z)$ and assume $T_K(z)\sim 10^4$ K. 

We see that in order to calculate the signal, we will need the redshift and space dependent functions: $n_{\rm HI}$, $v$, $dv/ds$, $T_{\rm K}$, $x_\alpha$ and $x_{\rm c}$. These will have to be provided by a simulation, which will generate boxes of each component for several redshifts. In our case, we used the output from the code \textsc{simfast21} as described later, but any other output can in principle be plugged in to our line of sight signal generator.

\subsection{Choosing the resolution}

The algorithm described above can be used to calculate the signal for any given observed frequency and angle. However, the observational maps are usually specified in a grid with a fixed angular and frequency resolution. The first question that then needs to be addressed is what resolution to choose for the map we want to construct as well as the optimal integration step $ds$ along the line of sight.

In terms of angular resolution, the optimal choice is a pixel with an angular size equal or slightly smaller than the angular size of the comoving simulation cells at the highest redshift we want to consider. This will guarantee that the signal is the same across the angular pixel (a smaller angular size could be used to take into account any sky curvature effects). In terms of frequency, again the pixel size should be related to the cell resolution of the simulation. A particular issue that might raise some confusion is whether such frequency resolution should take into account the line width, $\Delta\nu_{21}$. If we consider a single patch, $ds$, this will indeed generate an observed intensity that is only nonzero for a frequency interval of about $\Delta\nu_{21}/(1+z)$ (not factoring in peculiar velocities). This could imply that assuming the same intensity over the pixel for map pixels of larger sizes in frequency would be wrong. However, observed frequencies separated by more than $\Delta\nu_{21}/(1+z)$ will have contributions from patches along the line of sight which are very similar as long as the separation is smaller than the simulation cell resolution, meaning that indeed, it is the simulation cell resolution that needs to be used when choosing the map frequency resolution. In particular, we should take a pixel frequency size corresponding to the simulation cell size at its highest redshift.

Another issue in the implementation is whether the choice of the 21-cm line width makes a large impact in the final result since our algorithm indeed takes into account this value which can be subject to some uncertainty. In reality, the line width showing in $j_{21}$ from Eq. \ref{dI_main} will usually cancel with the frequency difference in Eq. \ref{ds_main}. First, as already seen, if the frequency change across $ds$ is larger and surround the line width, then $\nu_{\rm max}-\nu_{\rm min}=\Delta\nu_{21}$. On the other hand, when it is smaller than the line width, further steps down the line of sight will integrate Eq. \ref{ds_main} up to $\Delta\nu_{21}$ (assuming all other quantities to be constant) so that the width would again cancel out. This situation also explains when the line width becomes important: when it is large enough that the other quantities in the calculation are no longer constant. That is, when the HI patch contributing to a given frequency becomes larger than the simulation cell size. This is particularly relevant for regions with high optical depth.

As the relevant point is to make sure quantities are constant across the patch, for this implementation we choose the comoving integration step, $dx$ to be a fifth of the simulation comoving cell size. A final check is whether such size is consistent with the assumption of small optical depth across our small $ds=dx/(1+z)$ (which allows us to use equation \ref{dI_main}). For that, we need $k_{21} ds \ll 1$. Again, using typical values at $z\sim 10$ we get $k_{21} ds\sim 1\,{\rm K}/T_{\rm S}$ (with $ds = 0.05$ Mpc and a line width $\Delta\nu_{21}=60$ KHz). So, we should be always in the limit of low optical depth for practical values of the spin temperature. Note that this is true at each $ds$, not necessarily over the full integration over the line width. 

\section{Comparison with previous work}
\label{comparison}

The algorithm presented above takes a different perspective on the calculation of the 21-cm signal and it is therefore important to compare to what has been done before. First, due to the intrinsic radiative transfer process of the calculation, discussion of the 21-cm signal should go hand in hand with the light cone implementation. Our approach concentrates on the generation of the observed map. How to then resample this map in order to make it convenient for data analysis is a second stage process. The map itself can, for instance, have much finer resolution in the line of sight direction, or even have a non equally spaced grid. Other approaches tend to work directly on the simulation boxes and usually rely on cubic cells.

Besides our direct light cone/peculiar velocity coupling and map making process, our approach has two other main differences: First, we use the intensity from the previous iteration when integrating the radiative transfer equation, so multiple events can occur along the line of sight. This is specially relevant at high redshifts during absorption, which again, as far as we are aware, as not been dealt with before in terms of the proper RSD calculation. Second, we take into account the width of the 21-cm line, which as we will see below, can be relevant in certain situations.

\subsection{The standard approach}

Before we discuss some of the other methods that have been presented to address the peculiar velocities, we would like to first compare to what has been the standard way to deal with this. The basic assumption is the use of equation \ref{21-cm_standard}. There are then 3 situations where our approach will give different results:
\begin{enumerate}
\item
When $dv/ds\lesssim -H$. In this case the standard approach is to set a minimum threshold to $dv/ds$ to avoid the singularity (or negative temperatures), since this should only occur for high density, non-linear regions, which are expected to be scarce. However, as seen in Fig. \ref{pvhists}, the number of cells can be quite substantial. Capping the values in such a way, will generate high temperature pixels in the maps and have an impact on the power spectrum.
\item
When the assumption of thin optical depth is no longer valid, which is implicit in the standard equation. Our simulation naturally takes this into account by always using a small integration step $ds$ and using the output intensity from one event as input to the next event.
\item
When the 21-cm event "spreads" over a large enough region that the quantities are no longer constant (which would usually also mean a large optical depth). According to our calculation, we should only have 21-cm events across a redshift width of $\Delta z/(1+z) \sim 0.003$. This means that we should only expect differences from this effect if the simulation values change over this width. It also requires that the experiment has a frequency resolution better than this, which is usually the case.
\end{enumerate}

Our method, takes into account all these effects.
Another approach to the standard method would be to realize that, after calculating the values for each cell in the box with a given size $\Delta r$, translating to a frequency map would generate a pixel with frequency size:
\begin{equation}
\frac{\Delta\nu_0}{\nu_0}=\Delta r \frac{1+1/H \frac{dv}{ds}}{cH^{-1}}.
\end{equation} 
This would generate a map with pixels with different widths in the frequency direction and a rebinning to a common grid with appropriate weights would have to be done. This is the motivation for the method we discuss next.

\subsection{Cell shuffling}

Other methods have been proposed to deal with the peculiar velocity effects and avoid the singularity described above. Several schemes for constructing redshift space boxes from simulation data can be found in \citet{mao12}. Although it is referred there that the method can be applied to the arbitrary $T_S$ case, the algorithms described assume $T_S\gg T_{\rm CMB}$. Moreover, the input intensity is always assumed to be the CMB one, so that effectively, multiple scattering events are neglected. In this situation, we can imagine each HI atom as an emitter and the 21-cm signal is basically proportional to the HI number density in redshift space. The scheme that is more readily comparable is the Mesh-to-Mesh Real-to-Redshift-Space-Mapping scheme, e.g. MM-RRM as described in \citet{mao12}. Note again that these methods take the approach of going through the simulation box and calculating the signal instead of moving directly to the map making process as we do.

Similar versions of the MM-RRM scheme have been used in \citet{mellema06b}, \citet{jensen13} and \citet{Datta2014, Datta2012}. We focus here on the method presented in \citet{jensen13}. We assume that the code has produced a series of real space brightness temperature boxes, each representing the same comoving volume of the Universe at different cosmological redshifts. These boxes are assumed to consist of $N$ cells along the line of sight. Their algorithm is as follows:
\begin{enumerate}
\item Divide each cell along the line of sight into $n$ sub-cells, with each sub-cell assigned a brightness temperature $1/n'th$ of the original cell value.
\item Interpolate velocity and density fields onto this subgrid.
\item Shift the brightness temperature sub-cells according to the relation between real and redshift space: $s = r+\frac{1+z_{obs}}{H(z_{obs})}v_{||}(t,r)\hat{r}$, where $v_{||}$ is the peculiar velocity component along the line of sight.
\item Regrid the brightness temperature box to original resolution.
\end{enumerate}

This algorithm is then carried out for each output box separately. \citet{mao12} takes a slightly different approach in that the boundaries of each subgrid cell are shifted around instead of just the center. In this way, it allows for stretching as we consider in our case. Moreover, when regridding onto the regular, redshift-space space grid, only the overlapping part of the stretched cell is added. The approach in \citet{jensen13} is then equivalent to assuming that the sub-grid cells are so small that stretching can be neglected and that they fall completely in one or the other redshift-space cells. A somewhat non-trivial issue that seems to be ignored in these schemes is the intrinsic width of the 21-cm line, which implies that a give HI patch can contribute to different frequencies. 

In order to construct a light cone, the standard algorithm as described in Section \ref{sec:lc} is then used. However, contrary to our map making approach, in this case all points from the same redshift box are assumed to contribute to the same frequency. In reality, there may be events which occur at different cosmological redshifts but are observed at the same frequency due to the peculiar velocity effects. The two stage process (peculiar velocity+light cone) is used in \citet{Datta2014} where it is pointed out that the light cone correction needs to be made before the peculiar velocity correction.

\citet{Mondal2017} also consider the building of a light cone including the effects of peculiar velocities. Their algorithm go particle by particle within the simulation, assigning an observed frequency including the effects of peculiar velocities. Their approach splits the simulation into thin shells, which can be an issue for large peculiar velocities as it will not be assigned as actually being observed as if produced in another shell. In addition, they only consider the signal in emission at low redshifts when the spin temperature is much larger than the CMB.

In summary, we avoid the traditional two-step process, instead correctly treating the two corrections as inter-dependent, naturally  allowing for large optical depths and absorptions at higher z. 

Our algorithm presents a number of improvements compared to previous implementations: we allow for any value of $T_S$, we consider multiple scattering events (where the input intensity is the result of a previous event, which can be relevant during absorption) and we implicitly deal with events that go over a "wide" comoving length (or time interval), by doing the integration along the line of sight in fixed $ds$ steps. Moreover we calculate the line width from thermal broadening in each step and factor that in the 21-cm signal calculation.

\section{Simulation}
\label{simfast21}

\begin{figure}
\includegraphics[width=80mm]{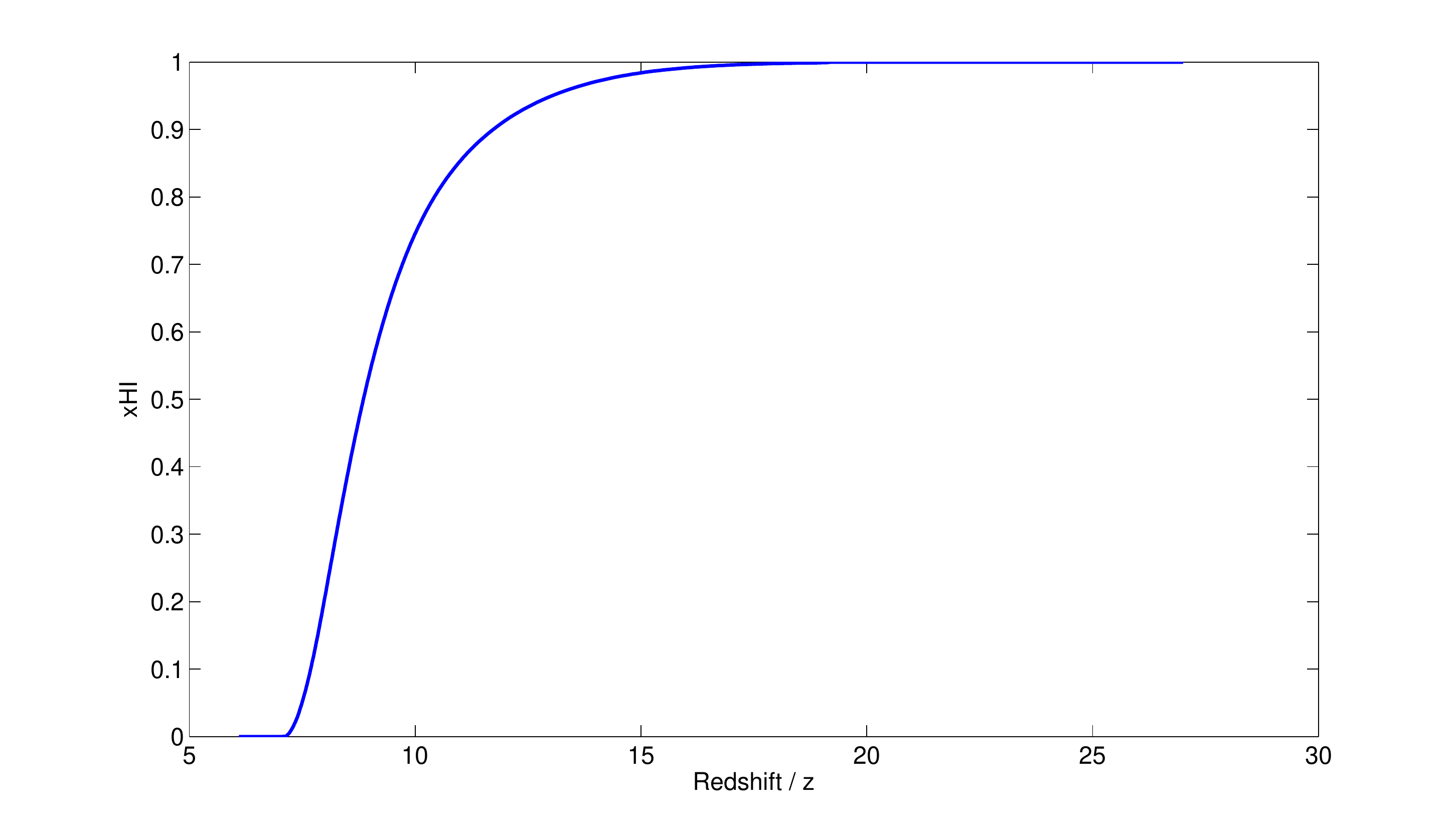}
\caption{The evolution of the neutral hydrogen fraction, $x_{\rm HI}$, with redshift in our simulation.}
\label{xHI}
\end{figure}

\begin{figure*}
\begin{minipage}{180mm}
\begin{center}
\includegraphics[trim={1.25cm 0 0 0},clip,width=55mm]{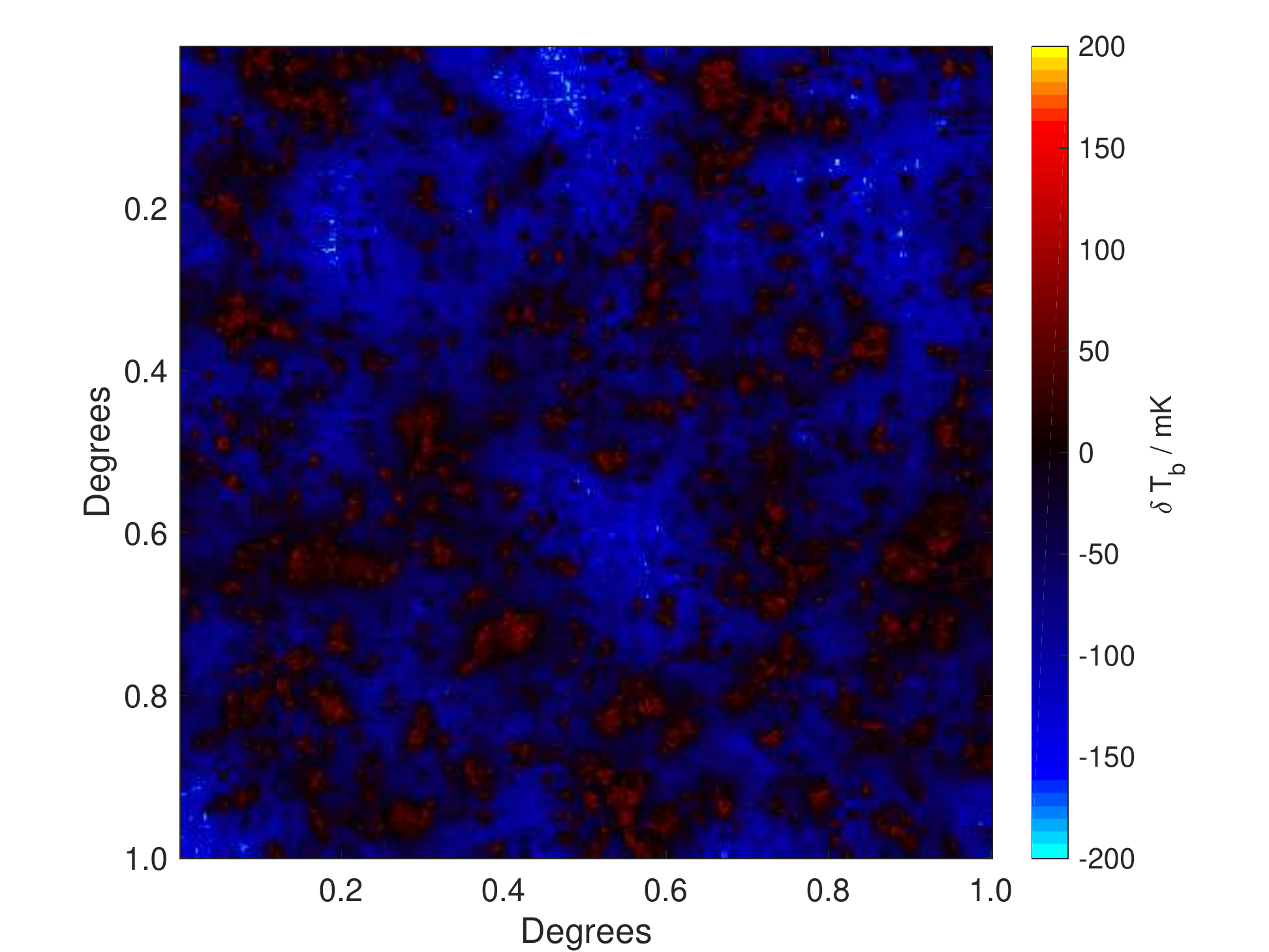}
\includegraphics[trim={1.25cm 0 0 0},clip,width=55mm]{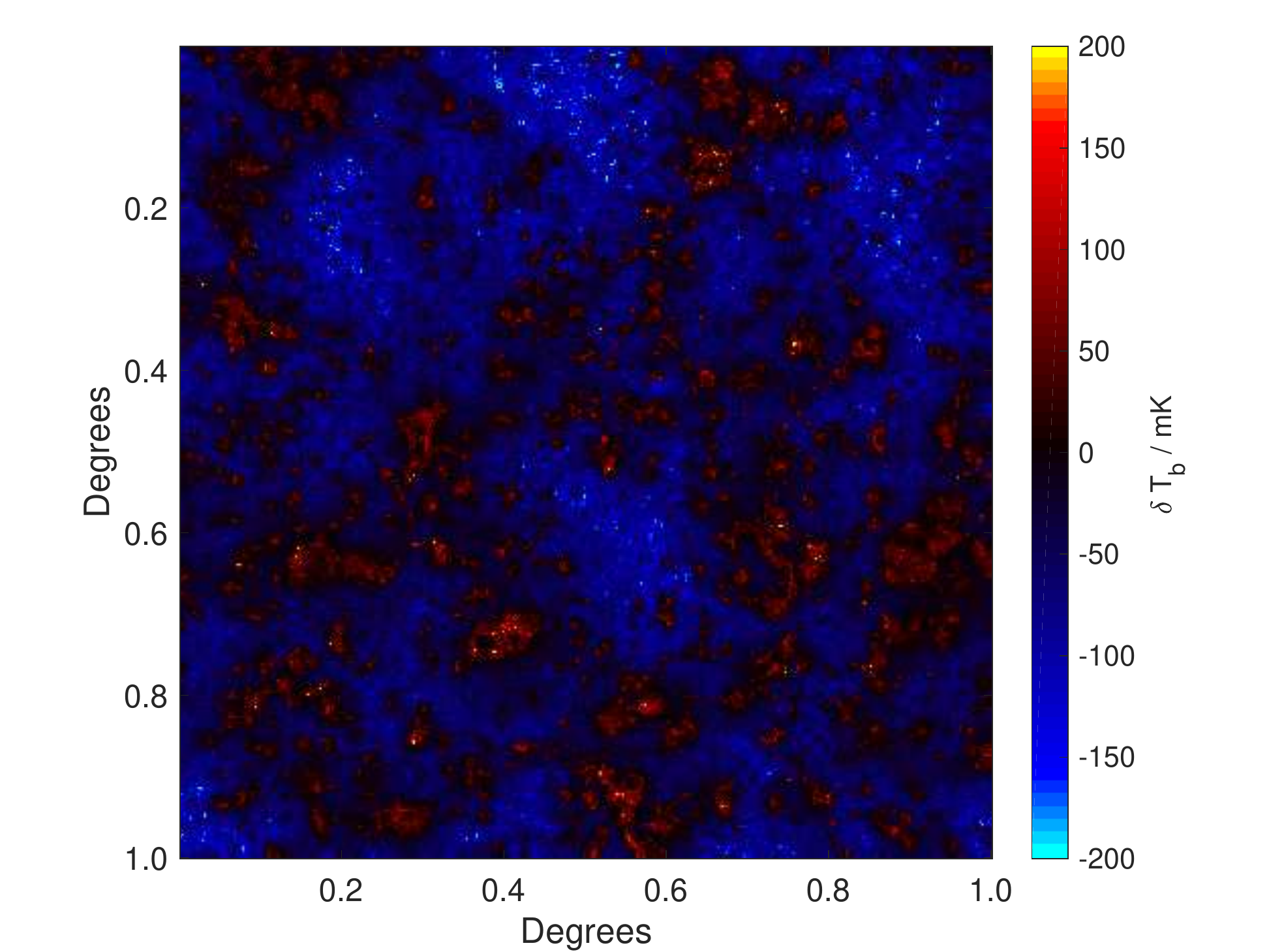}
\includegraphics[trim={1.25cm 0 0 0},clip,width=55mm]{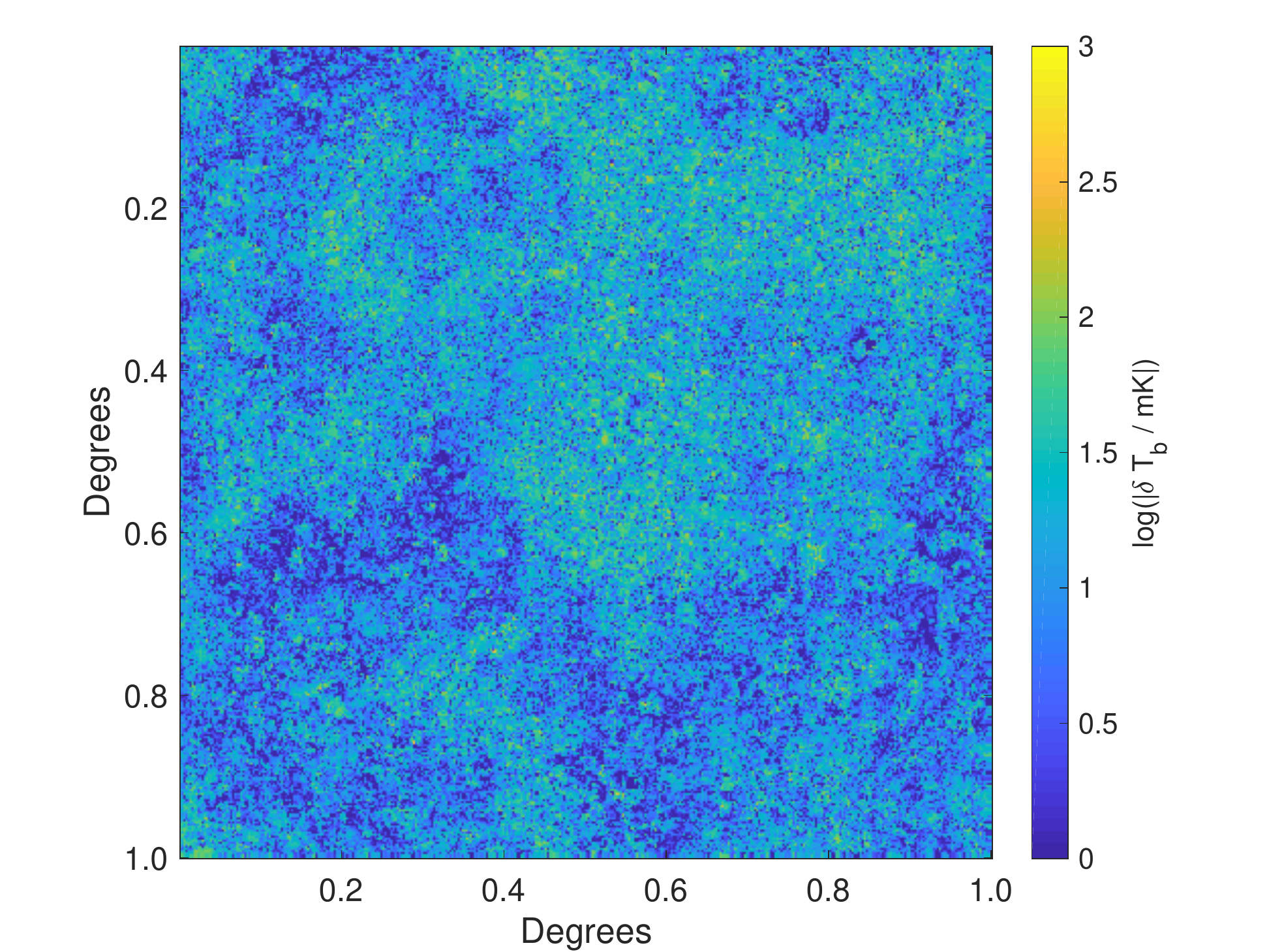}
\includegraphics[trim={1.25cm 0 0 0},clip,width=55mm]{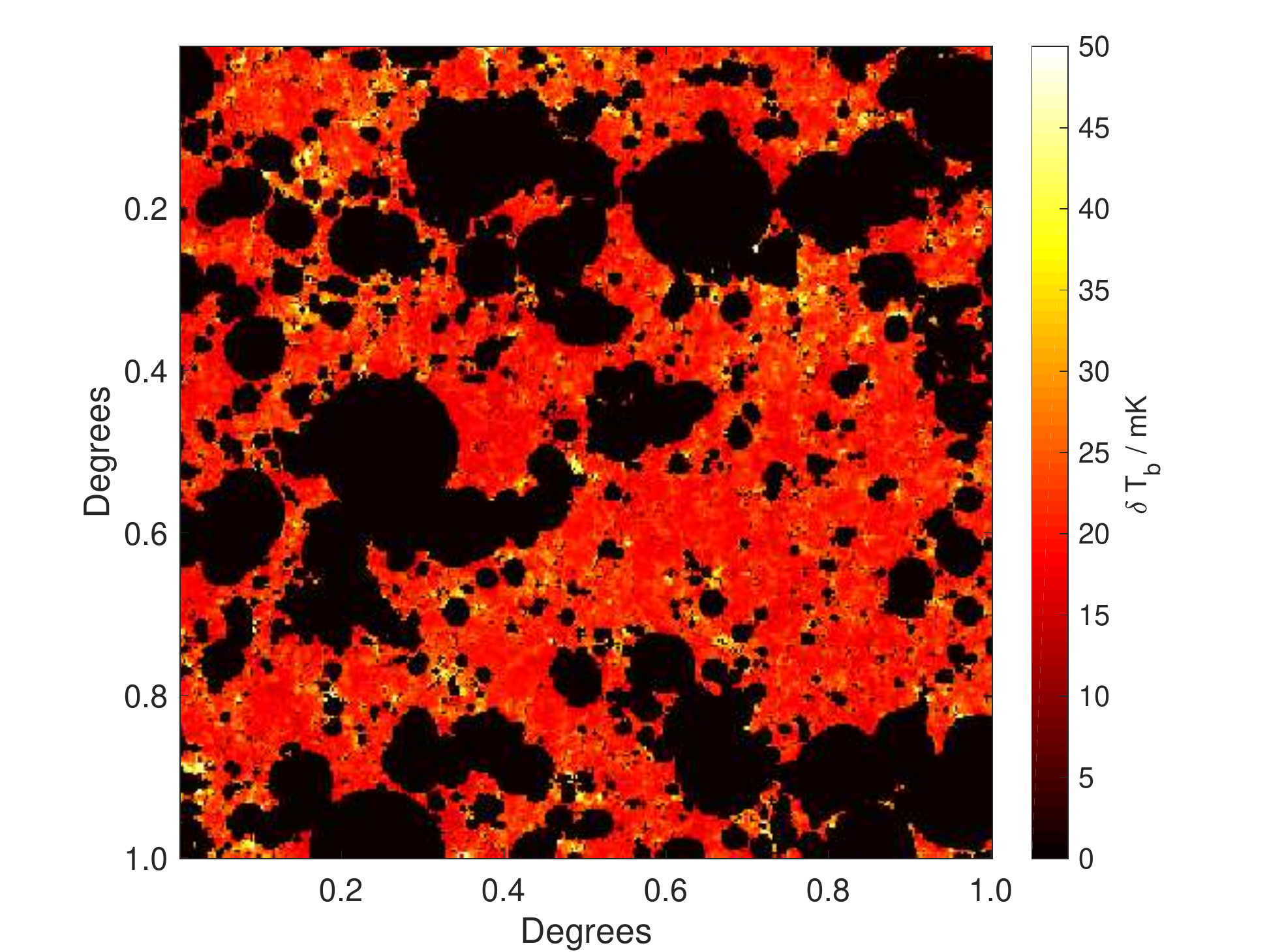}
\includegraphics[trim={1.25cm 0 0 0},clip,width=55mm]{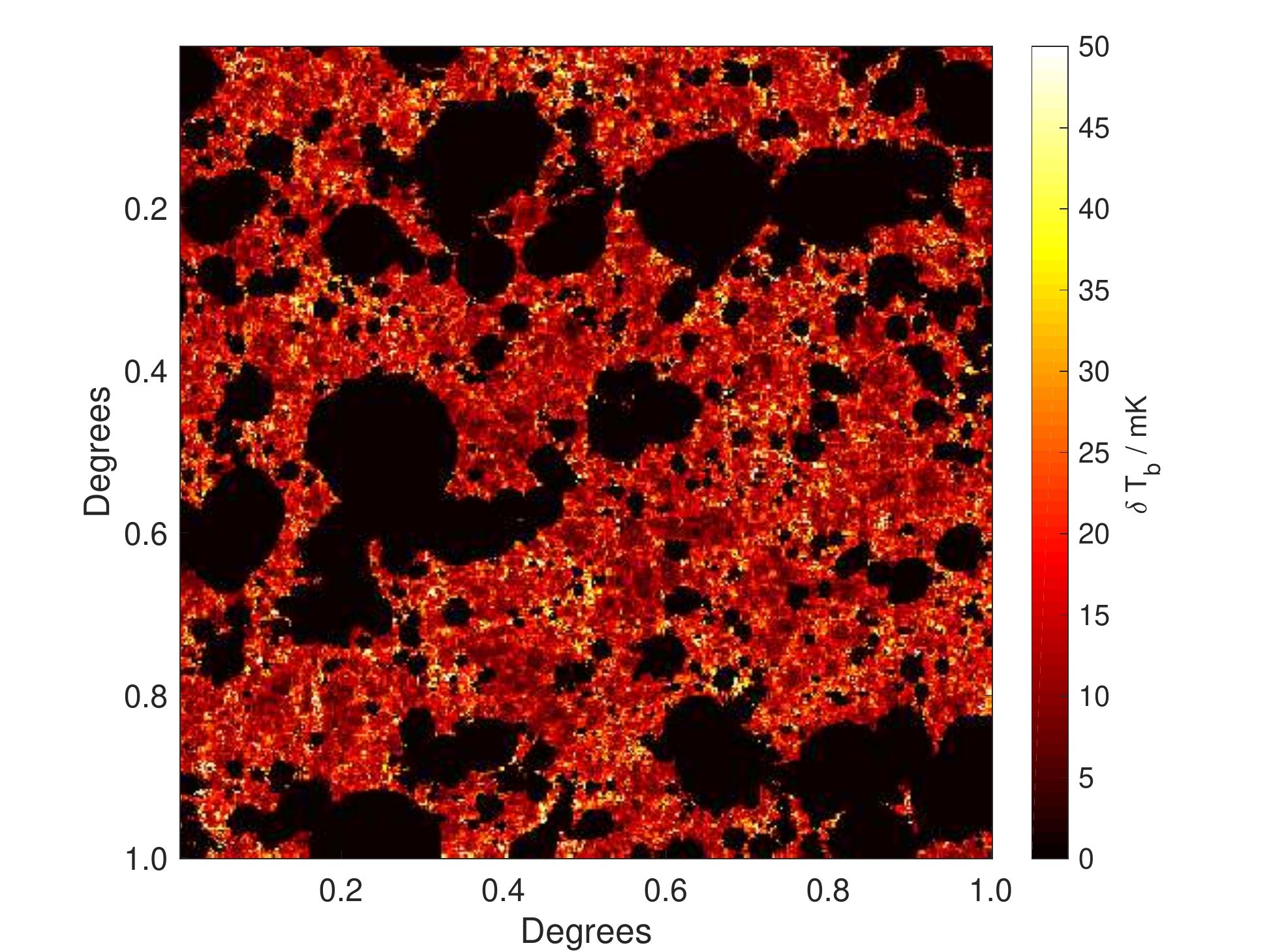}
\includegraphics[trim={1.25cm 0 0 0},clip,width=55mm]{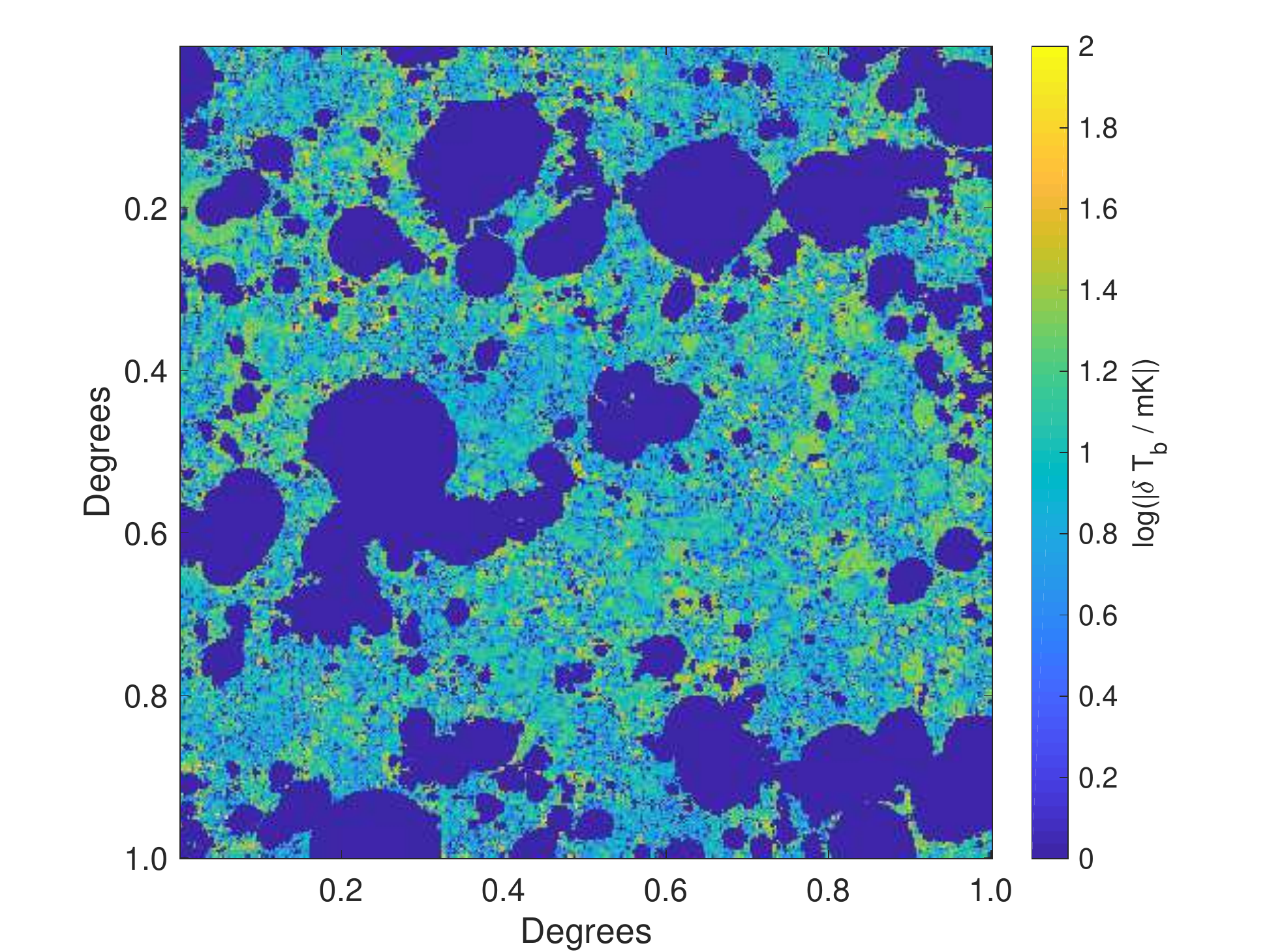}
\caption{Maps of the standard light cone (left), the new light cone (middle) and the absolute difference between these maps (|new - standard|) on the right. Top: 90 MHz (cosmic dawn/absorption). Bottom: 150 MHz (reionisation/emission).
Maps have a 1 degree field of view and a resolution of 9 arcseconds. While the structure appears similar on the largest scales on the sky, there is evidence of a different small-scale structure when using the standard light cone implementation and significantly different brightness temperature values on single pixel scales (note the log colorscale on the difference maps).}
 \label{mapsCD}
 \end{center}
 \end{minipage}
 \end{figure*}

\begin{figure}
\includegraphics[width=80mm]{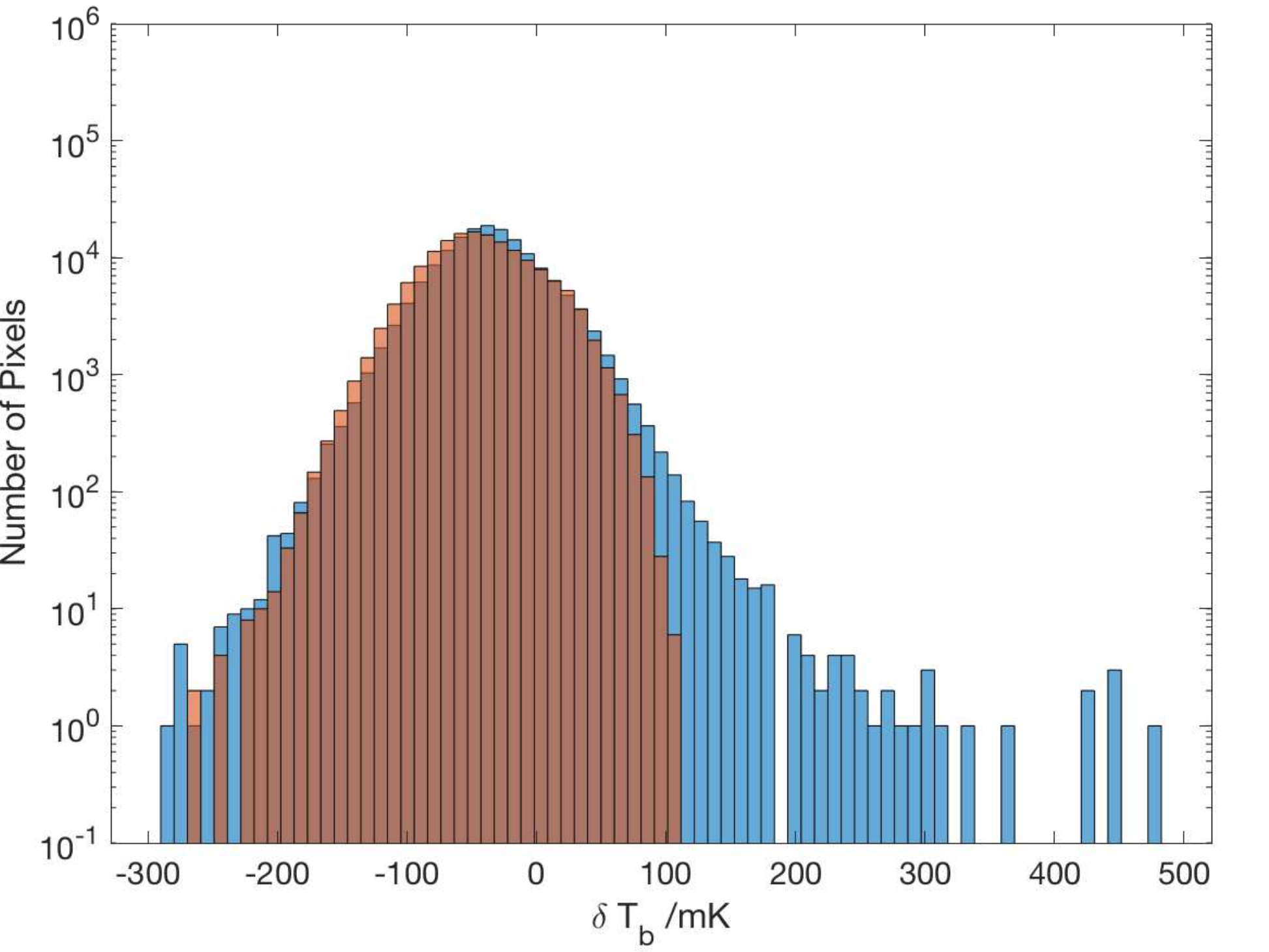}
\includegraphics[width=80mm]{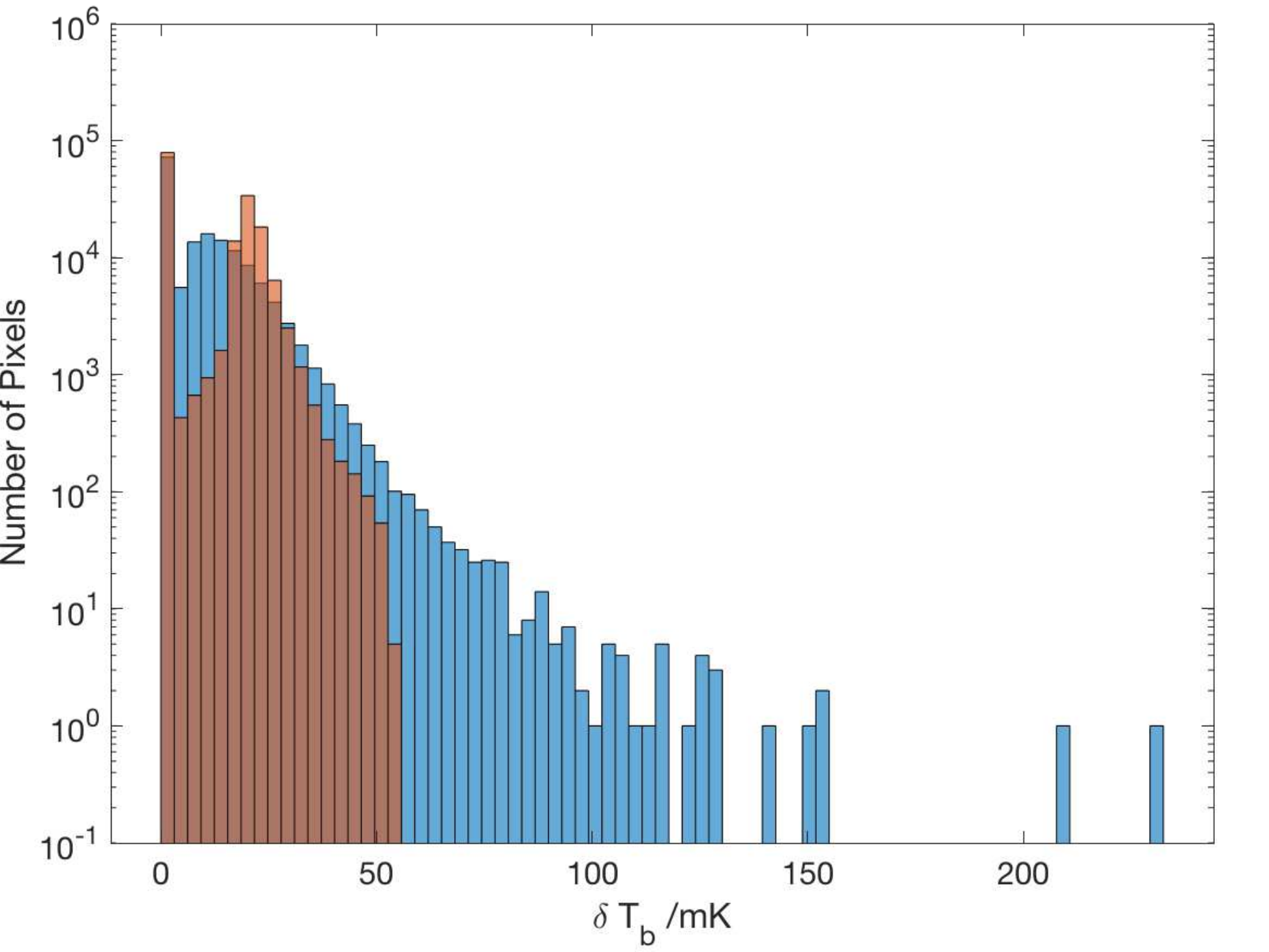}
\caption{Histograms of the standard light cone (red) and the new light cone (blue) for maps at frequencies 90 (top) and 150 MHz (bottom). The standard light cone method results in smaller tails of bright pixels, especially in emission. Note that the scale begins such that the lowest level bars represent single pixels. The large spikes at a brightness temperature of zero relate to ionized pixels. The new light cone method results in a distribution of pixel values with a longer bright tail due to the successful circumventing of a known issue in \textsc{simfast21} where when $dv/ds$ approached a value of $-H$, the brightness temperature diverged. This resulted in high temperature pixels being left out of the light cone.}
\label{hist}
\end{figure}

Though the new light cone code can
be applied to any simulation boxes, we apply it to \textsc{simfast21}\footnote{https://github.com/mariogrs/simfast21}. In response to the need for large field-of-view EoR simulations to
compare with observations, the semi-numeric scheme \textsc{simfast21}
was introduced by \cite{santos2010}. In the original simulation, while
peculiar velocity gradients were included in the calculation of the brightness
temperature, the effect on the observed light cone was not considered
and the simulation output was a real-space simulation box at each
redshift of interest. Our algorithm introduces a modification to this
code to output a light cone where the peculiar velocities are fully included. This modification replaces the original brightness temperature calculation, leaving all other parts of the code intact. For a full description of the original \textsc{simfast21} code, \citet{santos2010} should be consulted.

In the original \textsc{simfast21} implementation, after generating all the relevant boxes, the brightness
temperature would also be calculated at each redshift box, taking into account the peculiar velocity gradient:
\begin{eqnarray}
\delta T_b (z) &=& [23 \mathrm{mK}] x_{HI}(1+\delta(z)) \left( 1 -
  \frac{T_{CMB}(z)}{T_s}\right) \\ \nonumber
&& {} \times \left(\frac{h}{0.7}\right)^{-1} \left(\frac{\Omega_b h^2}{0.02}\right)
\left[\left(\frac{0.15}{\Omega_m
      h^2}\right)\left(\frac{1+z}{10}\right)\right]^{0.5} \\ \nonumber
&& {}  \times \frac{1}{1+ 1/H dv/dr} 
\end{eqnarray} 
As explained before, we see that the equation above breaks down when $dv/dr = -aH$. This can occur when either the
halo is virialized or at the turn around point of halo collapse just
before virialization.
To avoid singularities, \textsc{simfast21} defined a velocity gradient limit of $dv/dr > -0.7H$. Though this only affects a small number of cells, there has been evidence that this prescription could affect
more cells than necessary. \citet{mao12} found that significantly more
cells had  $dv/dr < -0.7H$ than were actually optically thick.
In the new brightness temperature code, we avoid the need for such
stark cut-offs completely by doing the integration along the line of sight for a fixed comoving $dx$.

We generated simulations to compare both approaches. We start by producing the quasi-linear density, velocity and ionization fields using the standard \textsc{simfast21} code. We output boxes every 0.5 in redshift between $z$ = 6 and 25. The initial
conditions are created on a $1200^3$ grid while a smoothed grid of
$400^3$ is used for the final boxes. All boxes are 200 Mpc per side and we choose an output light cone field of view of 1 square degree. If a field of view is equivalent to a distance larger than 200 Mpc at a certain redshift, the field of view is filled by assuming the boxes to be tiled on the sky. Since we expect the difference to show up on small scales and along the frequency direction, such field of view is adequate.

The evolution of the neutral fraction is shown in Fig. \ref{xHI}. 
We also show histograms of the peculiar velocities and peculiar
velocity gradients at a redshift of 8 in Fig. \ref{pvhists}. These
do not vary significantly over the redshift range. 

We then create a light cone of frequency resolution 0.1 MHz and 1 degree field of view with 9 arcsec resolution, for both the standard method and our new method. We would like to underline that in this paper we are always comparing our new results to the results using the standard light cone. This guarantees that we are comparing like with like since for instance the power spectrum without the light cone effect can be quite different.
Moreover, at each position along the line of sight $x(z)$, corresponding to a certain redshift $z$, the values we use are interpolated between the two redshift simulation boxes that bracket the redshift $z$ (taken at the same position in the box). We include this effect in the standard light cone method by interpolating the brightness temperature output from \textsc{simfast21} as we go down the line of sight. This way we guarantee that any difference in the two methods is only due to the new brightness temperature calculation algorithm. We describe below the differences that can be observed.

\section{Results}
\label{results}

\begin{figure}
\includegraphics[width=80mm]{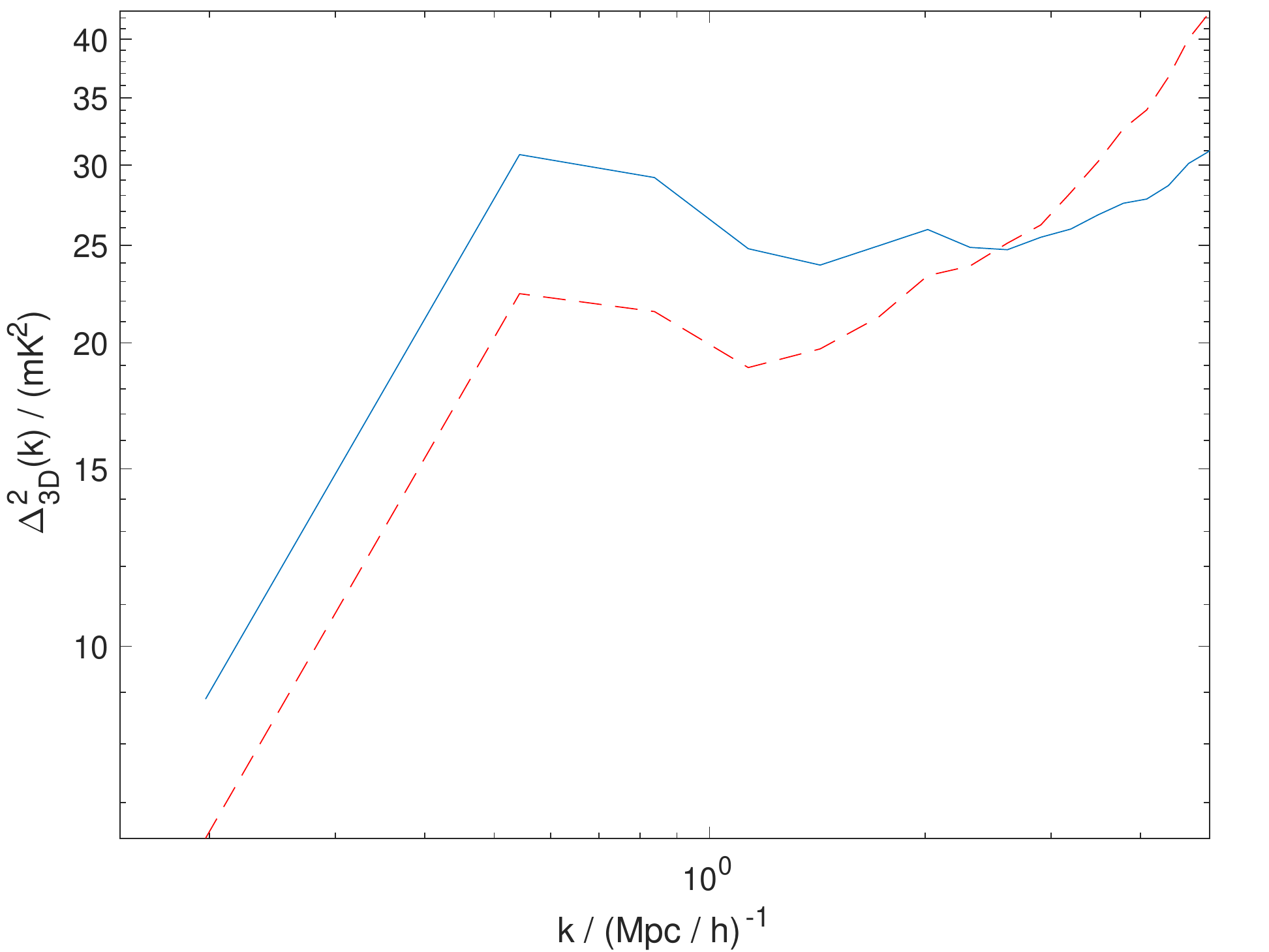}
\includegraphics[width=80mm]{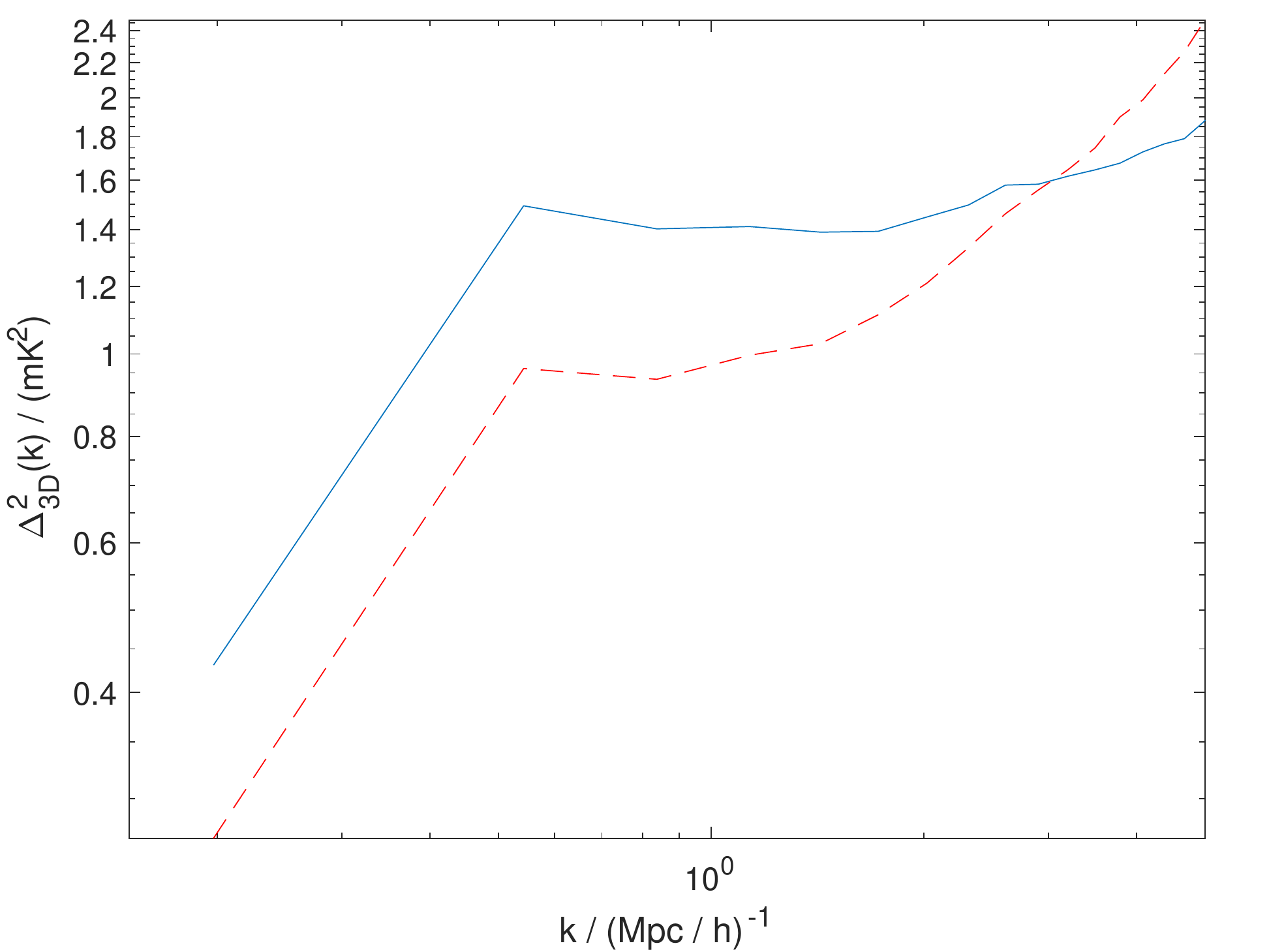}
\caption{In reading order we see 3D power spectra of the standard light cone (blue, solid) and the new light cone (red, dash) for 10 MHz bandwidths centered about frequencies 90 and 150 MHz.}
\label{fig:3Dps}
\end{figure}

We show the difference between the standard and new light cones within both the cosmic dawn (90 MHz) and EoR (150 MHz) in Fig. \ref{mapsCD}. Note that since we are only comparing 2D maps here, the standard light cone result is similar to just taking a slice of the coeval simulation box at a given redshift. The only difference will be if we have chosen a redshift between two coeval simulation boxes, in which case the slice would just be a linear interpolation between the two coeval boxes. When comparing the slices we see that the standard light cone maps appear smoothed in comparison to the new light cone maps. While the large scale structure is easily recognizable as similar between maps, the small scale structure is noticeably different, with evidence of significant differences in pixel brightness temperature values. This trend occurs across the frequency range tested.

To look at the difference in pixel values more closely, we plot the histogram of pixel values in the light cone maps for each frequency in Figure \ref{hist}. It is immediately clear that the new light cone method results in longer tails into bright temperatures, especially in emission. This is as a result of being able to consistently include high peculiar velocity gradient pixels in the new implementation. When $dv/ds$ approaches a value of $-H$, the brightness temperature diverges in the standard derivation of the brightness temperature. This was circumvented in previous implementations of \textsc{simfast21} by using a cut off preventing $dv/dr=-1$. The new light cone method naturally accounts for cells approaching this peculiar velocity gradient limit by considering an integration step along the line of sight as in Equation \ref{ds_main}. When $dv/ds=-H$ then $\nu_{max}-\nu_{min}=0$ and there is no divergence. In this case, $ds'=ds$ and, considering just the signal in emission, the intensity would be given by $j_{21} \Delta s$ (as in equation \ref{radeqn2}). Assuming this condition to be valid throughout the cell size (0.5 comoving Mpc in our simulation), then the maximum value in emission would be about 140 mK at $z\sim 10$ (well above the imposed cutoff in the old algorithm). This is consistent with what we see in the histograms, with a few cells in the box with larger values probably due to dark matter fluctuations (or the condition holding for more than 0.5 Mpc).

At 150 MHz and above we see reionization has begun by the lack of pixels showing 21-cm absorption and also the large peak at 0 mK representing ionized regions. The more compact nature of the brightness temperature pixel distribution, as a result of the cut off of these high peculiar velocity gradient pixels in the standard light cone implementation, results in the appearance of smoother light cone maps.

We calculate the 3D power spectrum of the new and standard light cones in order to compare how the simulations change with scale. The 3D power spectrum is defined as $\Delta_{3D}^2 = \frac{k^3 \mathrm{P}(k)}{2\pi^2 V}$ where V is the volume of the sub-band of the light cone being processed, $\mathrm{P}(k)=<\delta(k)\delta^*(k)>$ and $\delta(k)$ is calculated using the standard Fourier transform definition. The average is calculated over spherical shells in $k$. We choose sub-bands of 10 MHz in frequency range for the 3D power spectra and all frequencies referred to in the context of a 3D power spectrum refer to the central frequency of that sub-band.

\begin{figure}
\includegraphics[width=80mm]{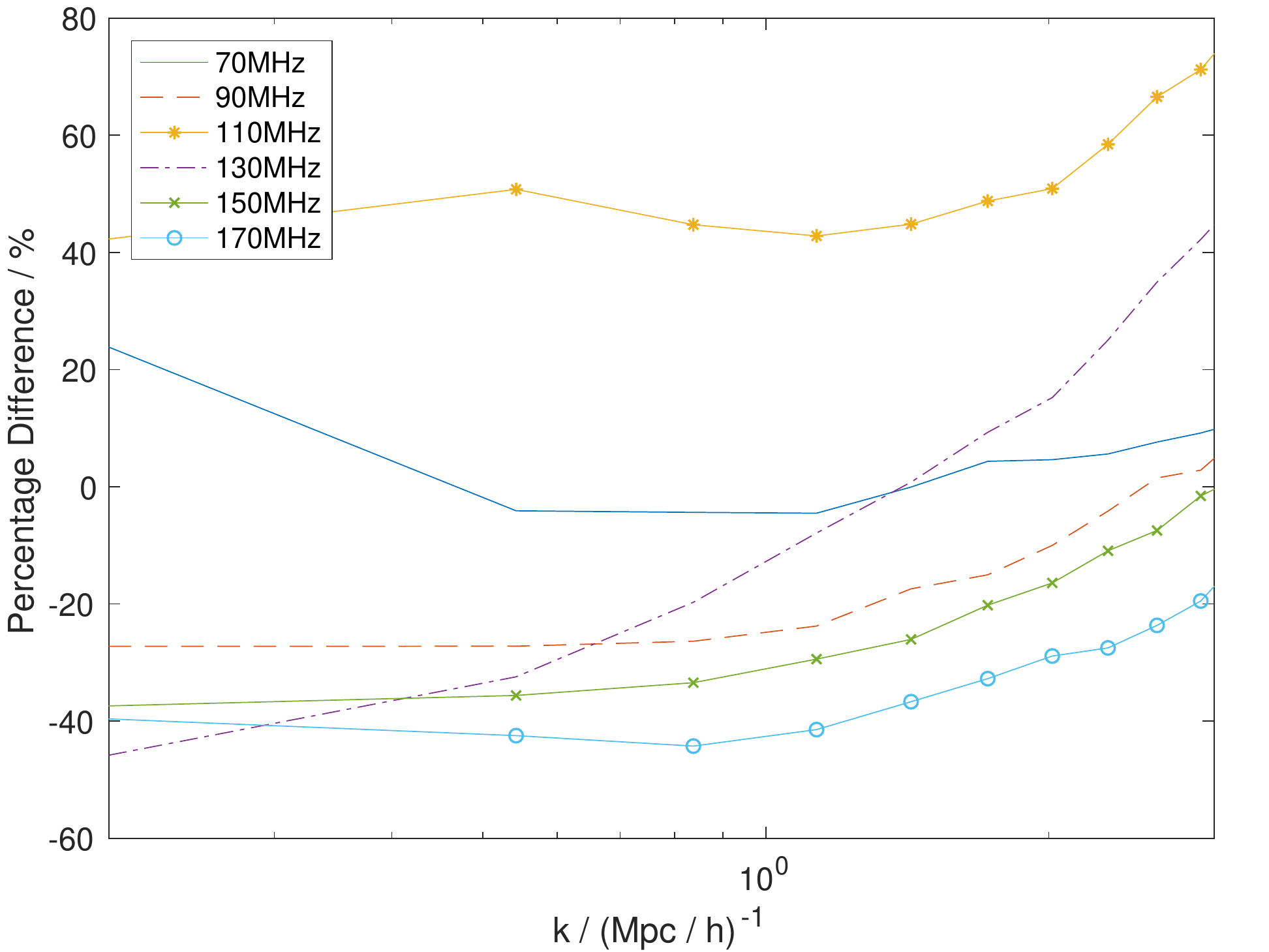}
\caption{The percentage difference of the 3D power spectrum between the new light cone and the standard light cone ($\frac{(\mathrm{P_{new}-P_{standard}})}{\mathrm{P_{standard}}} \times 100$) for frequencies 70, 90, 110, 130, 150 and 170 MHz. The new approach results in a significant change in power across frequencies and scales. There is a general trend of a drop in power at the largest scales on the sky and an increase in power at the smallest scales on the sky, for all frequencies apart from 110 MHz.}
\label{3DPS_ratio}
\end{figure}

In Figure \ref{fig:3Dps} we show the power spectra at 90 and 150 MHz and in Figure \ref{3DPS_ratio} we plot the percentage difference between the new and old light cone ($\frac{(\mathrm{P_{new}-P_{standard}})}{\mathrm{P_{standard}}} \times 100$) for a larger range of frequencies. Though there has been some discussion on how the 3D power spectrum risks omitting important information on the EoR \citep{Mondal2017,trott_lc_16} we include it here for completeness. 
The overall trend is that the difference increases towards larger $k$ (smaller scales). This is consistent with having a larger distribution of pixel values for the new algorithm as seen in the histograms. The extra pixels with large values are also expected to increase shot noise and therefore the power on smaller scales. From this plot it is clear that at various scales we see power losses of up to $45\%$ and power gains of up to 70\% when implementing the new light cone approach. This is a significant change in the power spectrum. 

\begin{figure}
\includegraphics[width=80mm]{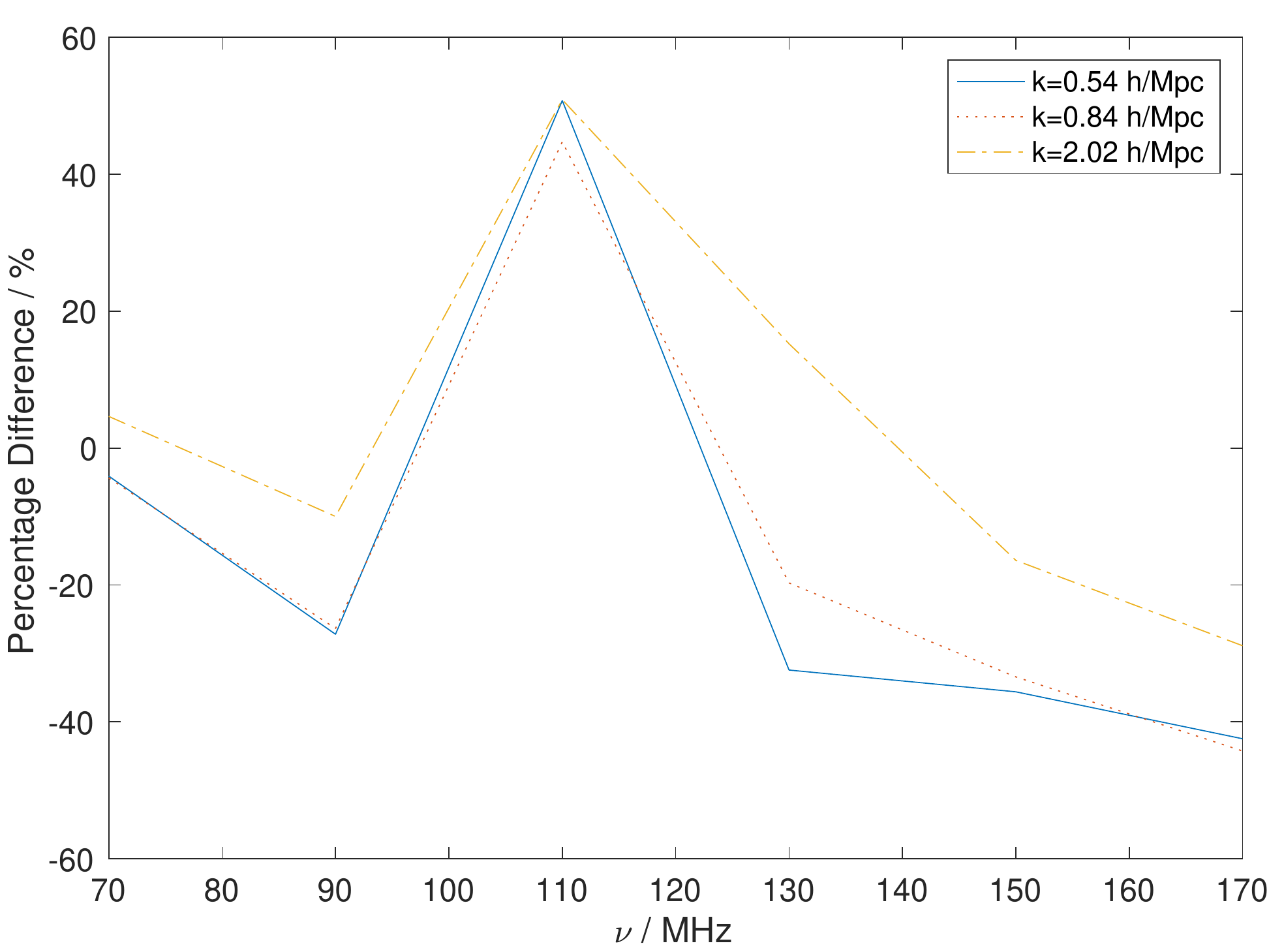}
\caption{Percentage difference evolution of the 3D power spectrum with frequency. The peak at 110 MHz is clearly seen, which we believe is connected to the beginning of reionization.}
\label{3DPS_evoln}
\end{figure}

The evolution of this percentage difference with frequency is plotted in Fig. \ref{3DPS_evoln}, highlighting the change in evolution which occurs at a peak at 110 MHz. Comparing to the 110 MHz histogram, we also saw a large difference in the pixel distribution similar to that shown for the 150 MHz histogram shown in Figure \ref{hist}, with a clear maximum value cutoff for the old method. This difference is probably due to the fact that we are at the beginning of reionization, when the the non-linearities become more important but are still in the neutral phase (e.g. not set to zero), so the effect of our algorithm when $dv/ds\sim -H$ is more obvious, allowing for a lot more higher pixel values overall at all scales. We emphasize again that these power spectra are calculated from the standard lightcone or the new light cone output, not simulation boxes at a fixed $z$.

\begin{figure*}
\begin{minipage}{180mm}
\begin{center}
\includegraphics[width=160mm,trim={1.4cm 14cm 1.4cm 11cm},clip]{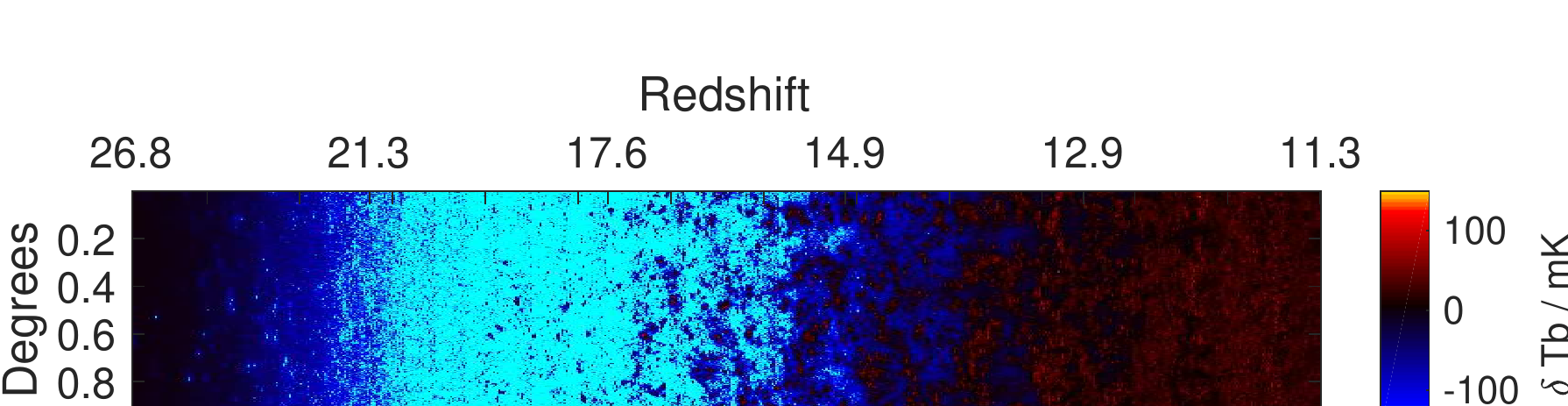}
\includegraphics[width=160mm,trim={1.4cm 12cm 1.3cm 13.1cm},clip]{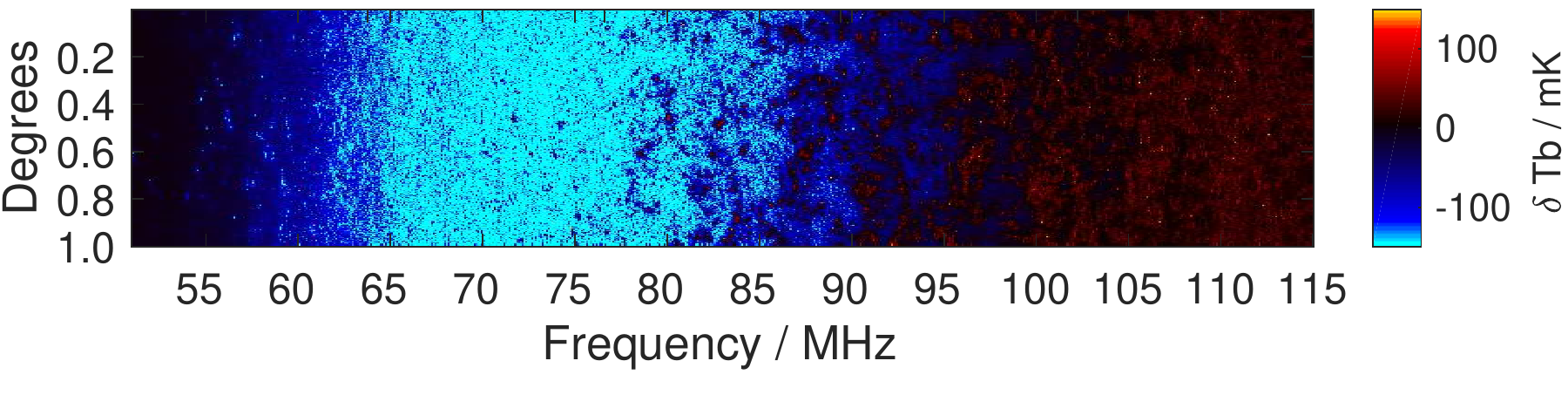}
\includegraphics[width=160mm,trim={1.4cm 14cm 1.5cm 11cm},clip]{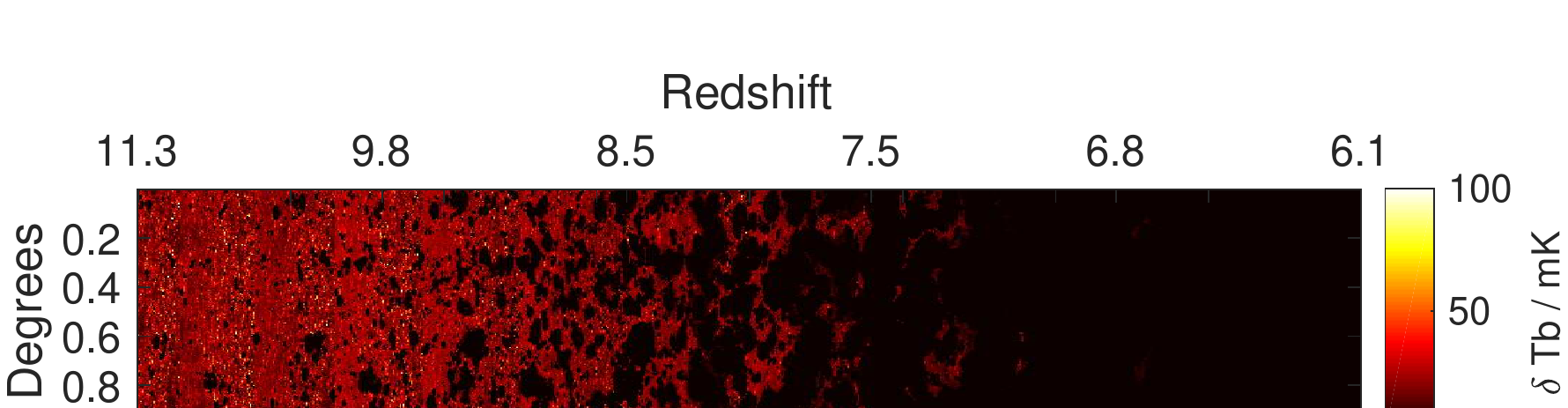}
\includegraphics[width=160mm,trim={1.4cm 12cm 1.5cm 13cm},clip]{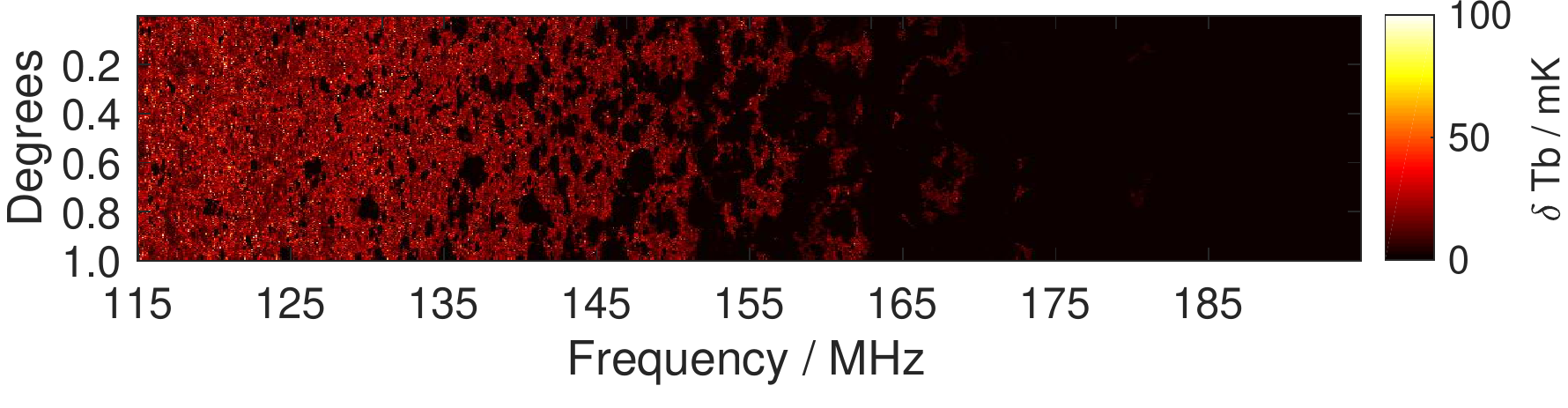}
\caption{A line-of-sight slice along the standard
  light cone (1st and 3rd) and the new light cone (2nd and 4th) during cosmic dawn (top two) and reionization (bottom two), where the line of sight is along the $x$ axis.}
\label{EoR_LOS}
\end{center}
\end{minipage}
\end{figure*}

We plot light cone line-of-sight slices along frequency in Figure \ref{EoR_LOS}. It is
evident that the general way in which reionization progresses is largely
unaffected by the new light cone. This is not surprising since the 
21-cm events at an observed frequency are not expected to come from
widely different redshift boxes. Moreover, the differences are harder to pick up here due to the x-axis resolution and the dynamic range of the color scale. In any case, we do see stronger differences at the onset of reionization as previously detected. These differences are more easily seen with standard statistics such as the cylindrical power spectrum which we analyse next.

\begin{figure*}
 \begin{minipage}{180mm}
 \begin{center}
 \includegraphics[width=140mm,trim={0.5cm 7cm 1cm 7cm},clip]{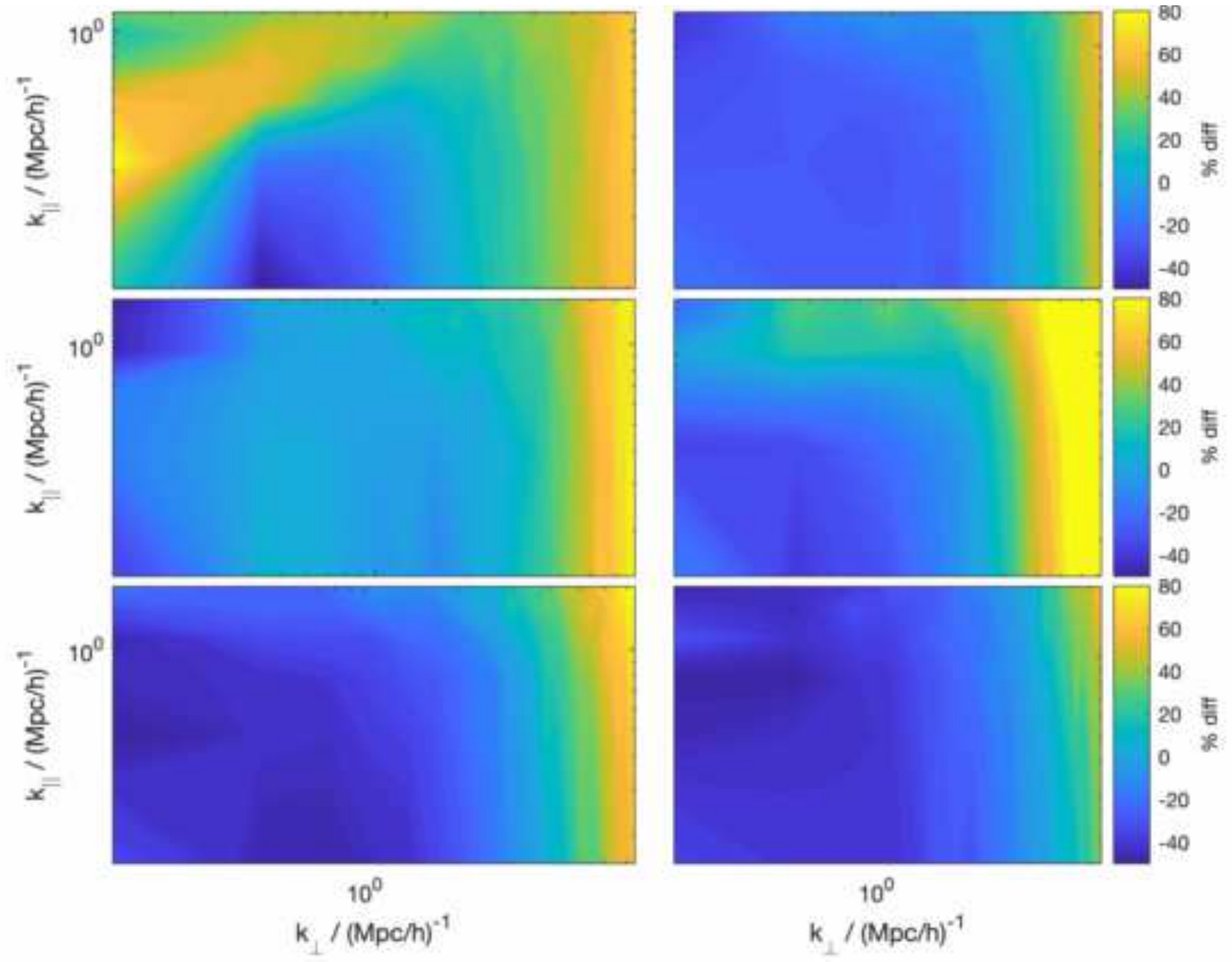}
 \caption{In reading order we see the percentage difference of the cylindrical power spectrum of a 10 MHz section of the standard light cone and the new light cone centered at frequencies 70, 90, 110, 130, 150, 170 MHz. $\mathrm{\% diff(k_{\perp},k_{\parallel})} = \frac{\mathrm{P_{new}(k_{\perp},k_{\parallel})-P_{standard}(k_{\perp},k_{\parallel}))}}{\mathrm{P_{standard}(k_{\perp},k_{\parallel})}} \times 100$.}
 \label{cylPS}
 \end{center}
 \end{minipage}
 \end{figure*}

A 2D cylindrical power spectrum in $(k_{\perp}, k_{\parallel})$ allows us to more easily
separate out the line-of-sight effects of the peculiar velocities. Such separation will be crucial to isolate the different astrophysical components from the cosmology as discussed in \citet{barkana05,santos2010}.
The power spectrum is calculated by 3D Fourier transforming the 10 MHz light cone volume
and binning the pixels according to both the perpendicular $k_{\perp}$
and parallel $k_{\parallel}$ Fourier scales.
 In Fig. \ref{cylPS} we show the percentage difference between the cylindrical power spectrum of the standard and new light cones. While the patterns of signal inflation and signal loss across the scales is consistent between the spherical and cylindrical power spectra, it is additionally apparent that these gains and losses are not symmetrical in $k$-space.
 
Anisotropies in the power spectrum are often measured in terms of
$(k,\mu)$ where $\mu$ is the angle with the line-of-sight, $\mu =
k_{\parallel}/|k|$. For an isotropic power spectrum we would expect $P_{\mu}(k,\mu)$
to be a flat line for a given $k$.

\begin{figure*}
\begin{minipage}{180mm}
\begin{center}
\includegraphics[width=80mm]{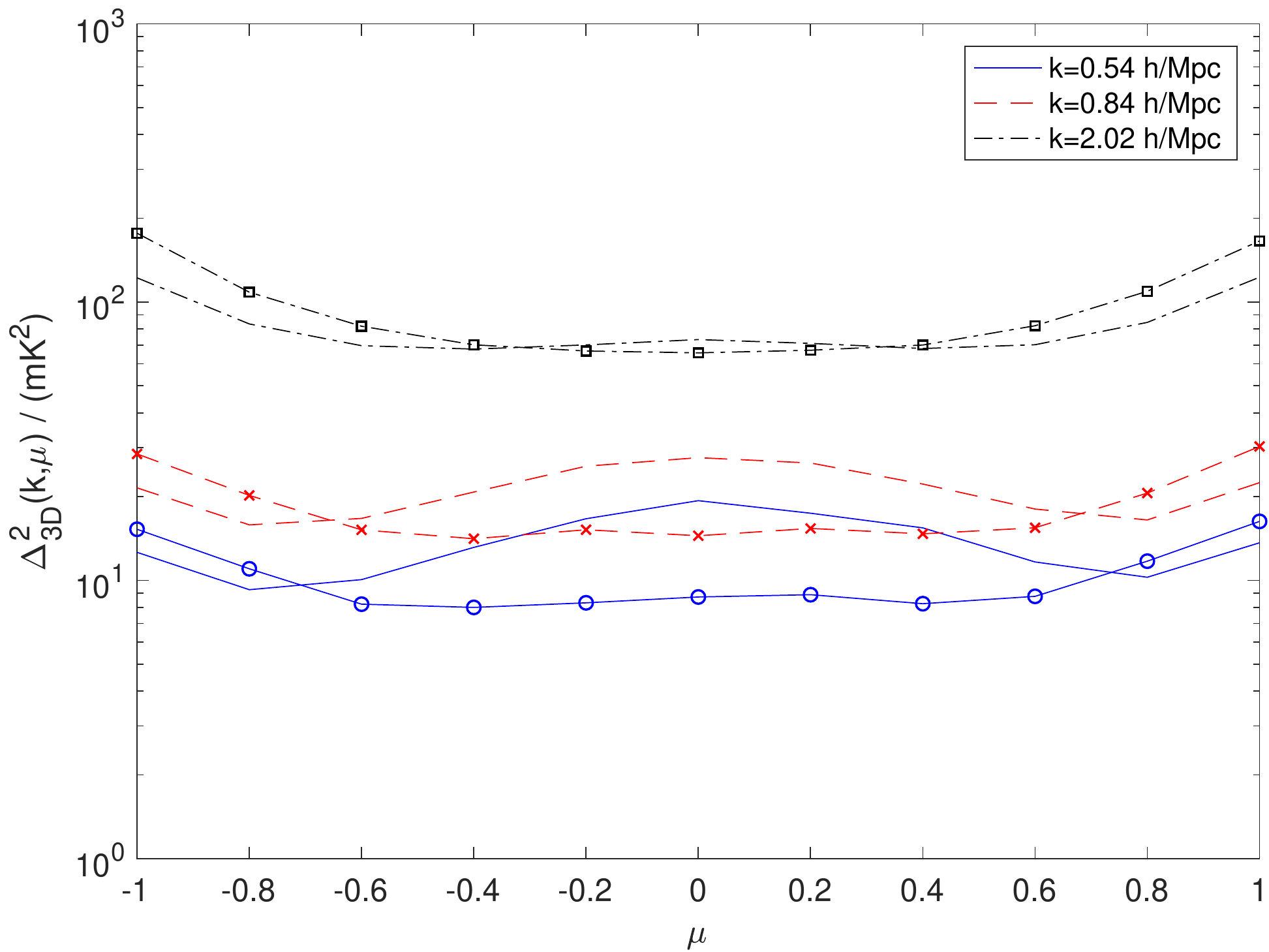}
\includegraphics[width=80mm]{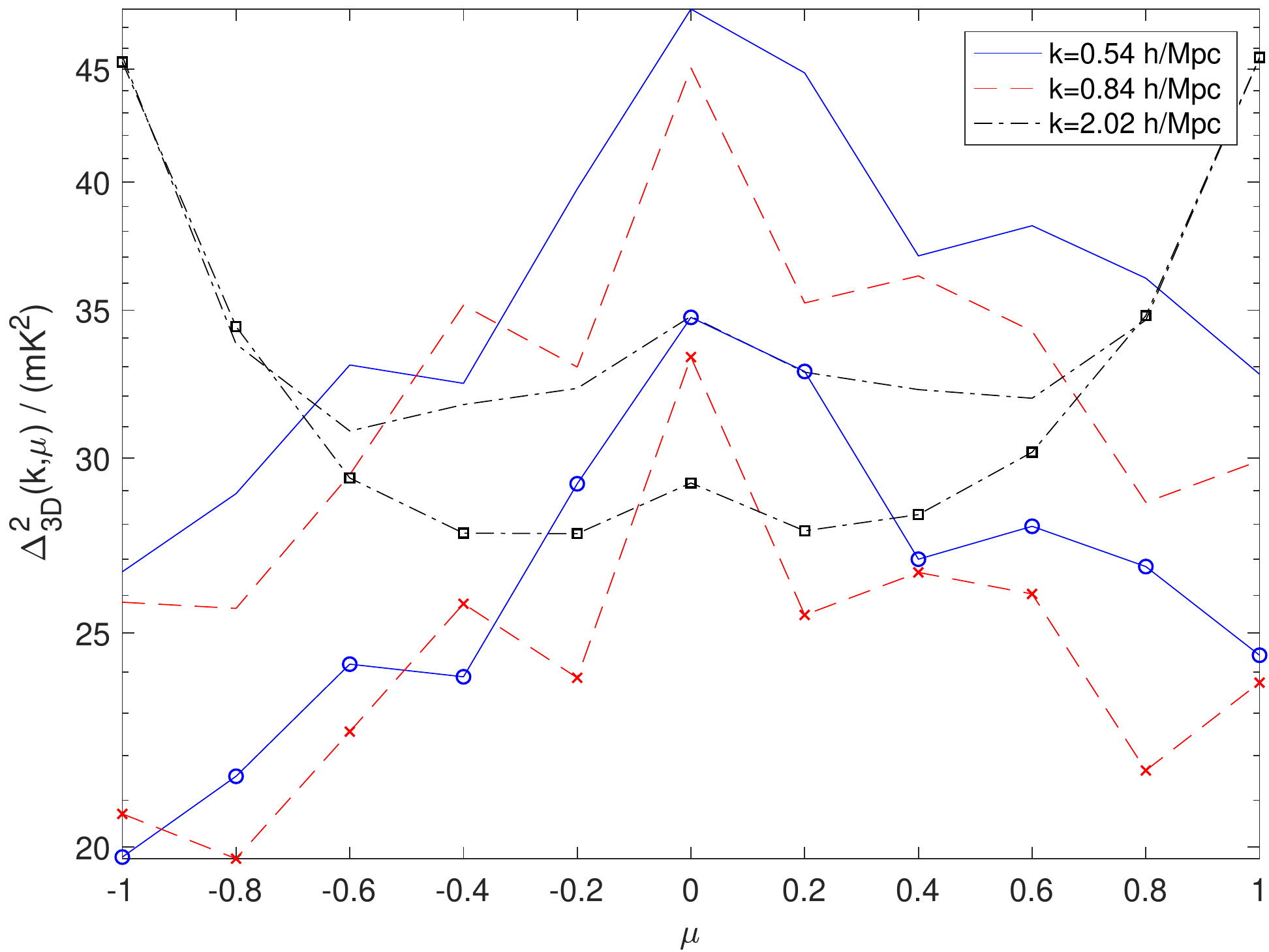}
\includegraphics[width=80mm]{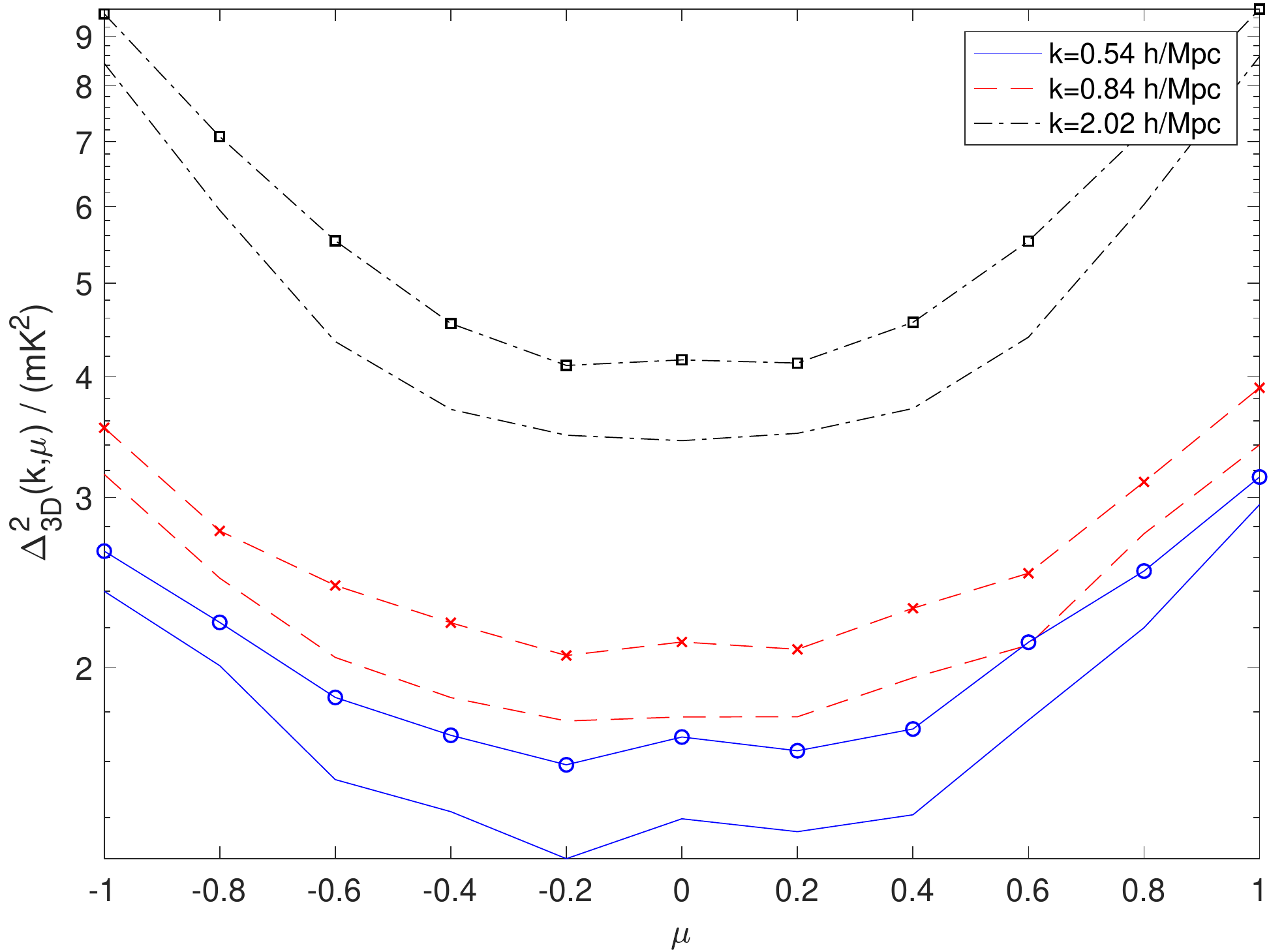}
\includegraphics[width=80mm]{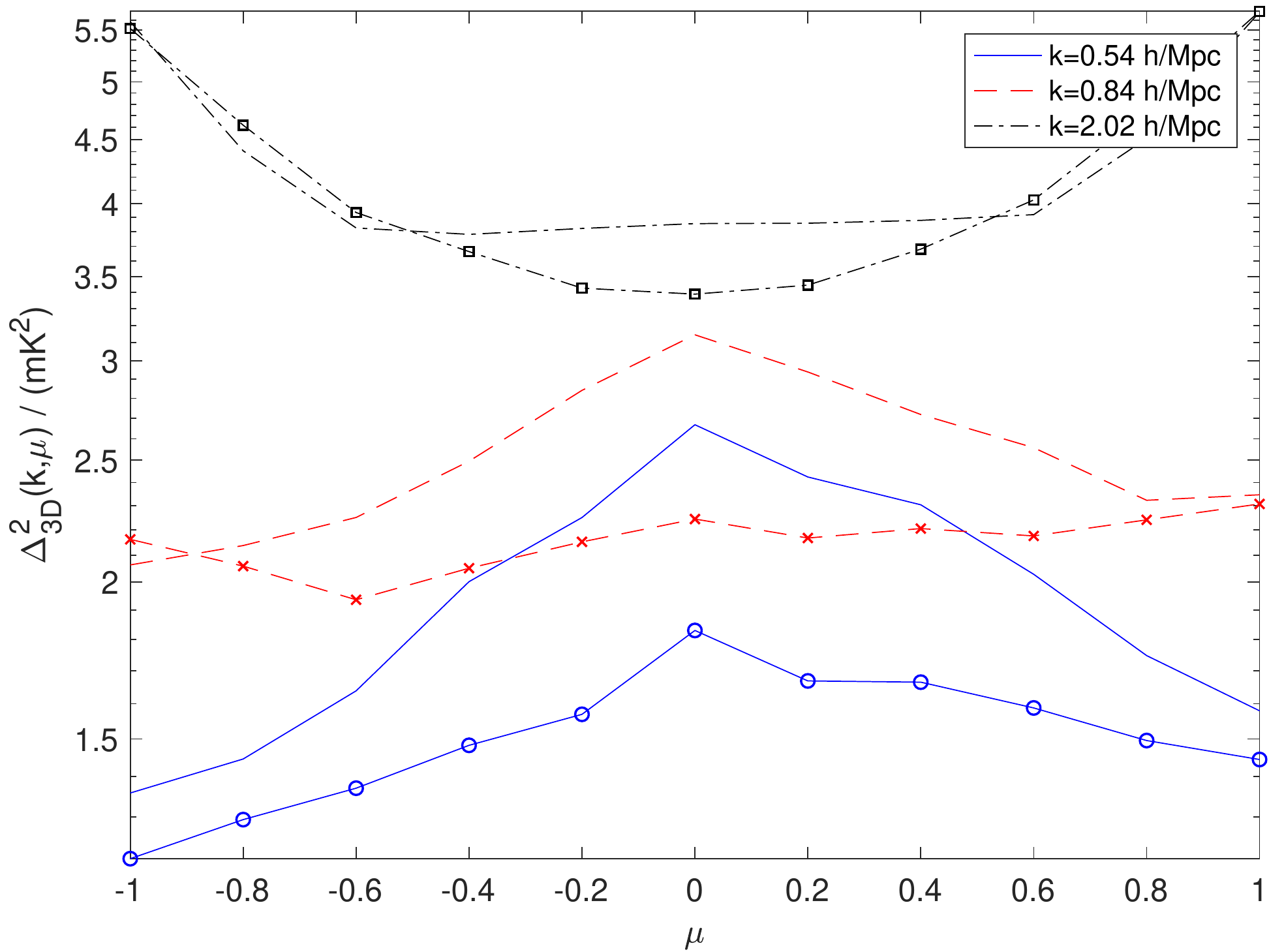}
\includegraphics[width=80mm]{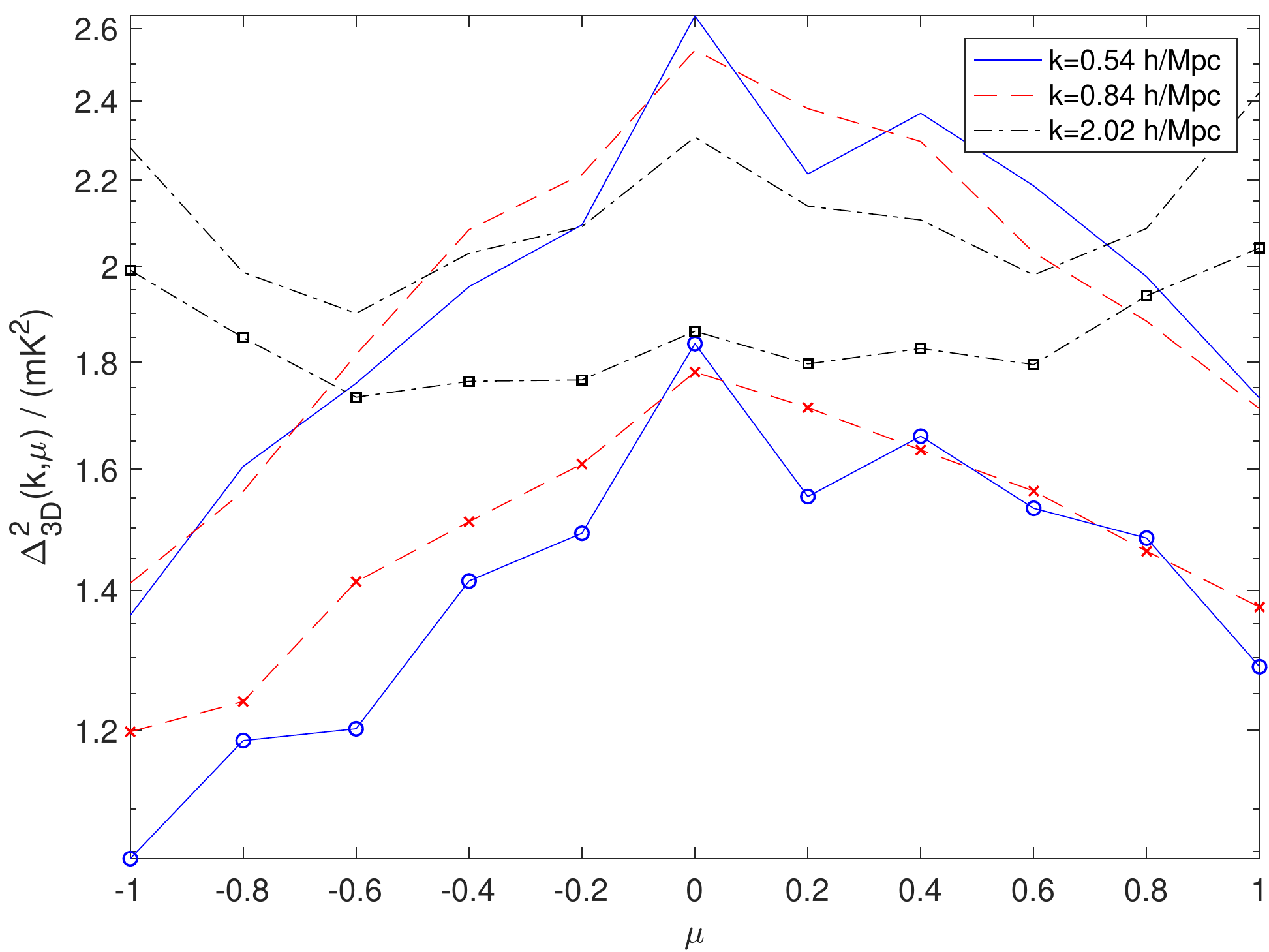}
\includegraphics[width=80mm]{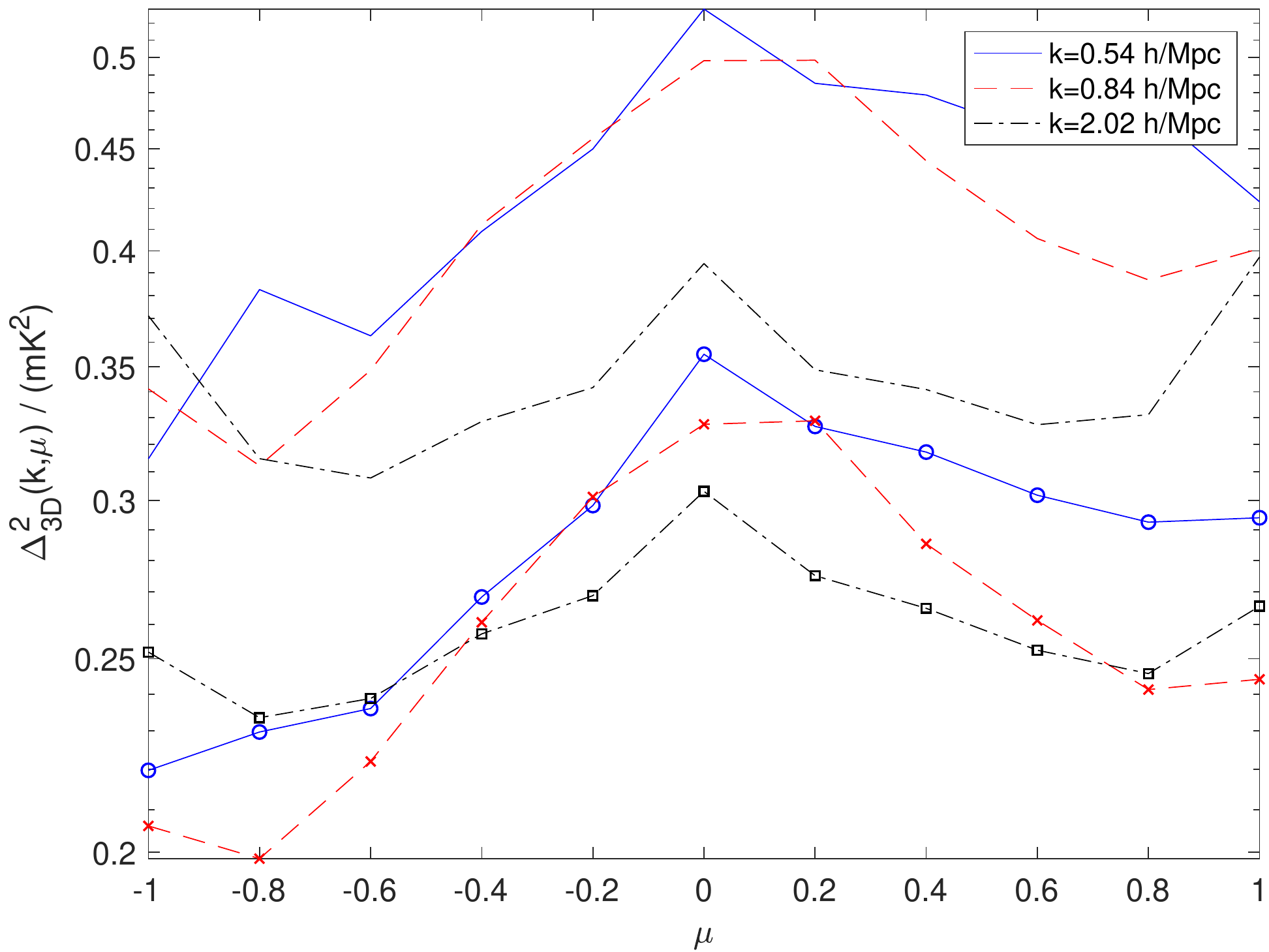}
\caption{In reading order: The $\mu$ decomposition of the 3D power spectrum for 10 MHz sections centered on 70, 90, 110, 130, 150 and 170 MHz, for the standard light cone (lines) and new light cone (lines+points).}
\label{mufig}
\end{center}
\end{minipage}
\end{figure*}

\begin{figure}
\includegraphics[width=80mm]{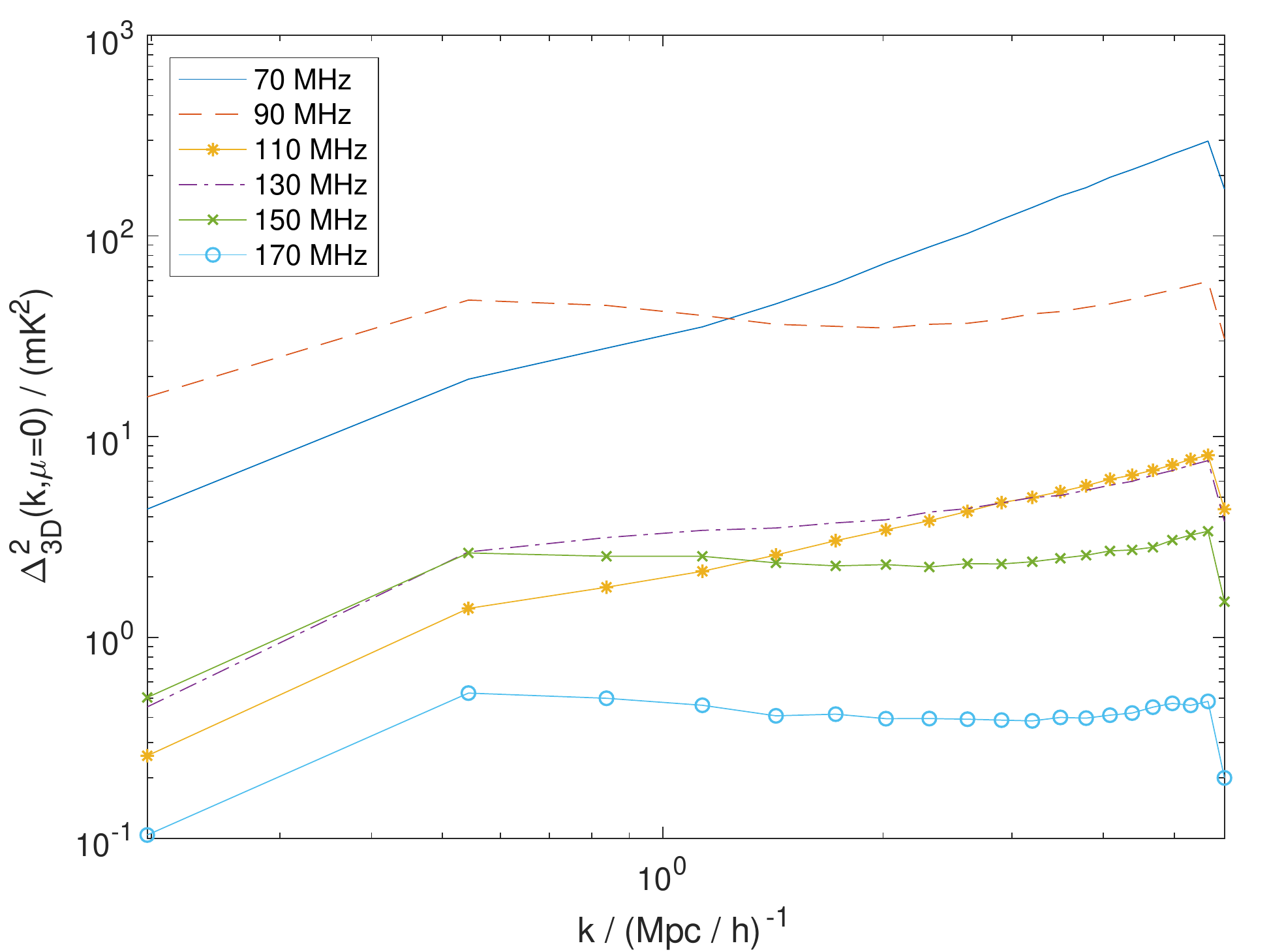}
\includegraphics[width=80mm]{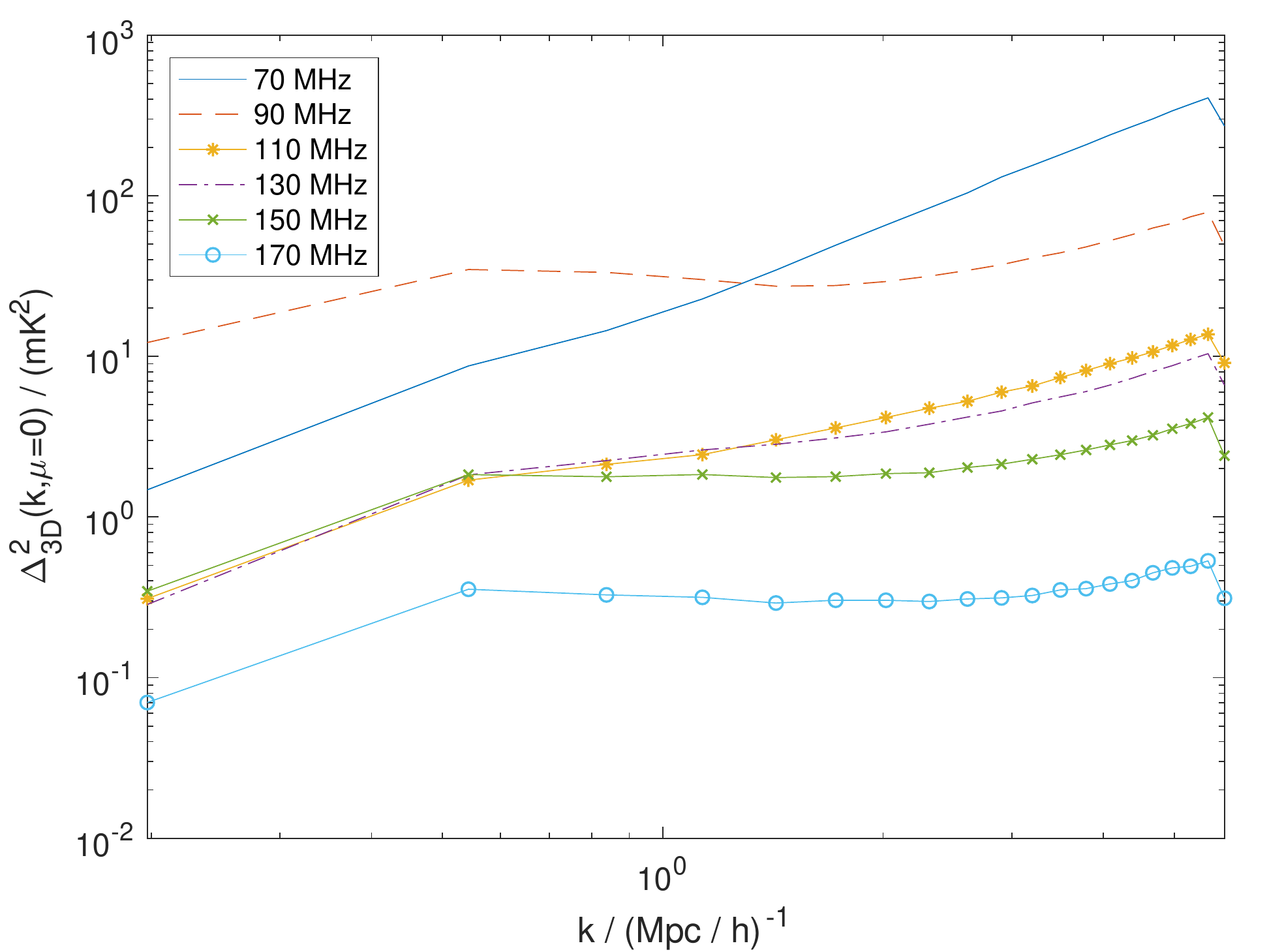}
\includegraphics[width=80mm]{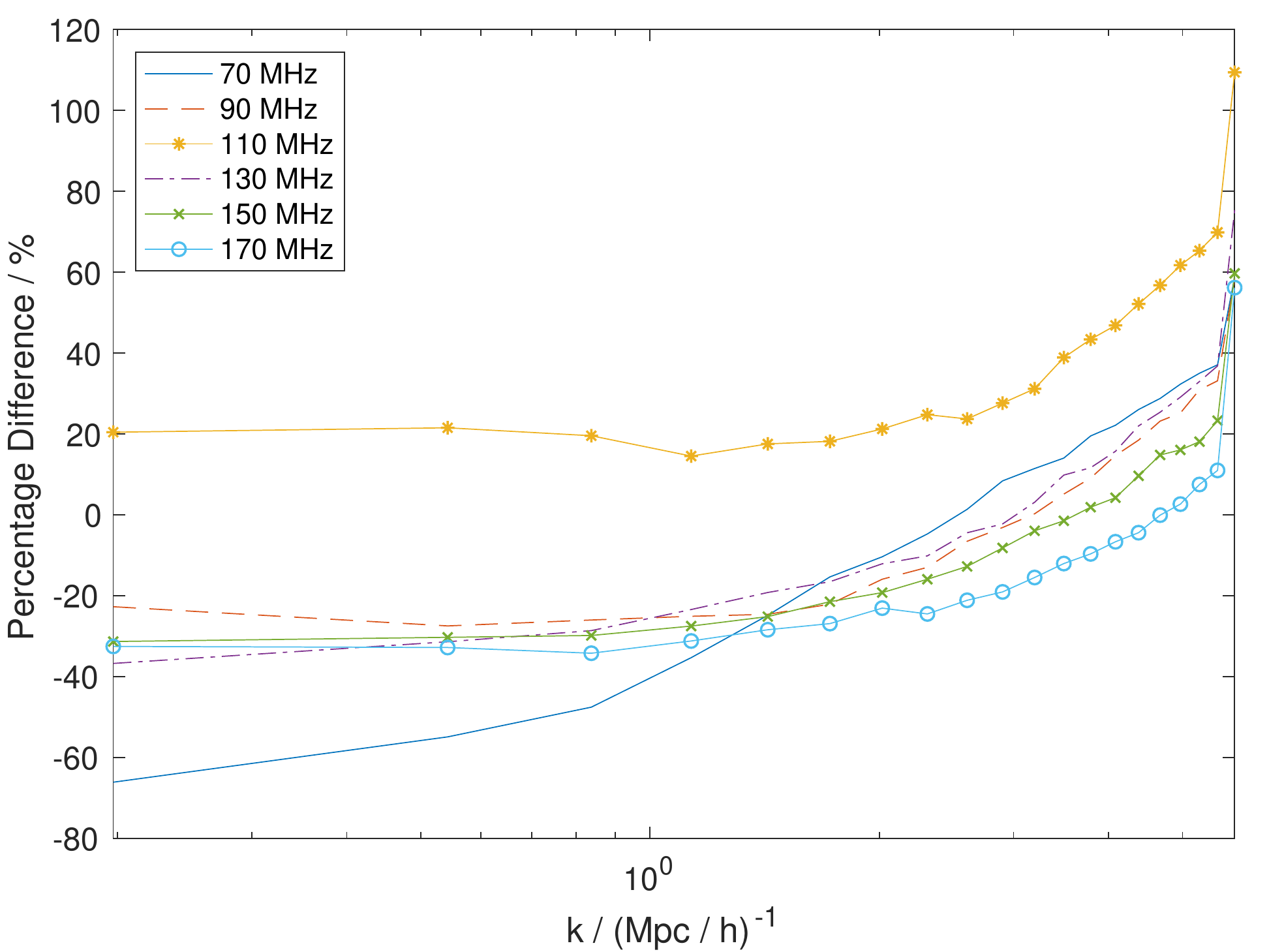}
\caption{From top to bottom: the evolution of the $\mu=0$ bin of the $\mu$-decomposed power spectrum, for the standard light cone, the new light cone and the percentage difference between the two.}
\label{mu_evoln}
\end{figure}

The standard and new light cone power spectra, $\Delta^2_{3D}(k,\mu)$, are shown in Fig. \ref{mufig}. We see power loss at almost all scales, with the exception of 110 MHz. This is consistent with what we have seen before since here we only consider scales $k<1.5$ h/Mpc.
The shape of the power spectrum remains the same except at 70 MHz and 150 MHz where we see larger differences in directions approaching $\mu=0$. To investigate this further we plot the $\mu=0$ bin only of the power spectrum for the standard and new light cones and the percentage difference between them in Figure \ref{mu_evoln}. We see that for most frequencies there is a drop in the power in the $\mu=0$ bin across all scales. For 70 MHz and 110 MHz we see a power increase on the largest $k$ scales. 

It is clear that any fitting of a $\mu$-decomposed polynomial
such as described in \citet{barkana05,santos2010} would result in differing
values for the cosmological and astrophysical power spectrum terms
motivating this full inclusion of peculiar velocities into the light cone.

\section{Conclusions}
We have presented a new open source code (\url{https://github.com/chapmanemma/RSD_LC_github}) which calculates the redshift space light cone from the standard real space simulation output of any reionization
code, while properly taking into account the coupling with peculiar velocities. We set out a physically motivated treatment of peculiar velocities, both in the 21-cm intensity equation and in the calculation of the observed
frequency for each 21-cm event. This requires the individual assessment
of each location along a line-of-sight according to whether a 21-cm event would be observed at
the frequency of observation and how the intensity must be adjusted as
a result of the interaction of cosmological redshift and the peculiar velocity. We have shown that this full treatment produces significant statistical differences when compared to the light cone assembled by taking slices from coeval boxes equivalent to the redshift-frequency conversion. 

Our algorithm naturally deals with high velocity gradients, large optical depths/multi-scattering events and is applied at low frequencies/high redshifts during absorption. It also factors in the line width in the calculation. We find that the new method increases the number of bright pixels present in the light cone due to the natural dispensing of the divergence in the brightness temperature equation, when $dv/ds = -H$. Previously, \textsc{simfast21} implemented a cut off on cells approaching these values, resulting in a loss of bright temperature cells in the light cone. With the new implementation we no longer lose this information. 

We find a general trend of power
loss with the application of the new light cone on the larger scales on the sky, while at the smallest scales we see a power increase. We see an overall power increase at 110 MHz (though with the same trend) which we believe is connected to the early stage of reionization.

We conclude that to be
confident of any analysis made from simulated light cone data, one must
be sure to fully incorporate the peculiar velocities as shown here. Imposing a hard cutoff on the peculiar velocities will dramatically change the pixel distribution as seen in the histograms.
In future work we plan to use a higher resolution simulation and multiple reionisation models in order to further assess when the difference in power spectrum, which results from using this full treatment of the light cone, is most significant. 


\section{Acknowledgments}
EC acknowledges the support of a Royal Astronomical Society Research Fellowship and studentship funding from the Science and Technology Facilities Council in the preliminary stages of this project. In later stages EC acknowledges the Royal Society Dorothy Hodgkin Fellowship.
MGS acknowledges support from the South African Square Kilometre Array Project and National Research Foundation (Grant No. 84156).

The authors would like to thank Garrelt Mellema, Hannes Jensen and Kanan Datta for useful discussions.

\bibliographystyle{mnras}  
\bibliography{peculiar_lcone}

\begin{thebibliography}{}
\makeatletter
\relax
\def\mn@urlcharsother{\let\do\@makeother \do\$\do\&\do\#\do\^\do\_\do\%\do\~}
\def\mn@doi{\begingroup\mn@urlcharsother \@ifnextchar [ {\mn@doi@}
  {\mn@doi@[]}}
\def\mn@doi@[#1]#2{\def\@tempa{#1}\ifx\@tempa\@empty \href
  {http://dx.doi.org/#2} {doi:#2}\else \href {http://dx.doi.org/#2} {#1}\fi
  \endgroup}
\def\mn@eprint#1#2{\mn@eprint@#1:#2::\@nil}
\def\mn@eprint@arXiv#1{\href {http://arxiv.org/abs/#1} {{\tt arXiv:#1}}}
\def\mn@eprint@dblp#1{\href {http://dblp.uni-trier.de/rec/bibtex/#1.xml}
  {dblp:#1}}
\def\mn@eprint@#1:#2:#3:#4\@nil{\def\@tempa {#1}\def\@tempb {#2}\def\@tempc
  {#3}\ifx \@tempc \@empty \let \@tempc \@tempb \let \@tempb \@tempa \fi \ifx
  \@tempb \@empty \def\@tempb {arXiv}\fi \@ifundefined
  {mn@eprint@\@tempb}{\@tempb:\@tempc}{\expandafter \expandafter \csname
  mn@eprint@\@tempb\endcsname \expandafter{\@tempc}}}

\bibitem[\protect\citeauthoryear{{Ba{\~n}ados} et~al.,}{{Ba{\~n}ados}
  et~al.}{2018}]{Banados2017}
{Ba{\~n}ados} E.,  et~al., 2018, \mn@doi [\nat] {10.1038/nature25180}, \href
  {https://ui.adsabs.harvard.edu/abs/2018Natur.553..473B} {553, 473}

\bibitem[\protect\citeauthoryear{{Barkana} \& {Loeb}}{{Barkana} \&
  {Loeb}}{2005}]{barkana05}
{Barkana} R.,  {Loeb} A.,  2005, \mn@doi [\apjl] {10.1086/430599}, \href
  {https://ui.adsabs.harvard.edu/abs/2005ApJ...624L..65B} {624, L65}

\bibitem[\protect\citeauthoryear{Barkana \& Loeb}{Barkana \&
  Loeb}{2006}]{Barkana2006}
Barkana R.,  Loeb A.,  2006, \mn@doi [MNRAS]
  {10.1111/j.1745-3933.2006.00222.x}, 372, L43

\bibitem[\protect\citeauthoryear{Becker, Bolton  \& Lidz}{Becker
  et~al.}{2015}]{Becker2015}
Becker G.~D.,  Bolton J.~S.,   Lidz A.,  2015, \mn@doi [Publ. Astron. Soc.
  Aust.] {10.1017/pasa.2015.45}, 32, e045

\bibitem[\protect\citeauthoryear{Bharadwaj \& Ali}{Bharadwaj \&
  Ali}{2004}]{Bharadwaj2004}
Bharadwaj S.,  Ali S.~S.,  2004, \mn@doi [MNRAS]
  {10.1111/j.1365-2966.2004.07907.x}, 352, 142

\bibitem[\protect\citeauthoryear{Bowman, Rogers, Monsalve, Mozdzen  \&
  Mahesh}{Bowman et~al.}{2018}]{Bowman2018}
Bowman J.~D.,  Rogers A.~E.,  Monsalve R.~A.,  Mozdzen T.~J.,   Mahesh N.,
  2018, \mn@doi [Nature] {10.1038/nature25792}, 555, 67

\bibitem[\protect\citeauthoryear{{Choudhury}, {Haehnelt}  \&
  {Regan}}{{Choudhury} et~al.}{2009}]{Choudhury2009}
{Choudhury} T.~R.,  {Haehnelt} M.~G.,   {Regan} J.,  2009, \mn@doi [\mnras]
  {10.1111/j.1365-2966.2008.14383.x}, \href
  {http://adsabs.harvard.edu/abs/2009MNRAS.394..960C} {394, 960}

\bibitem[\protect\citeauthoryear{{Ciardi}, {Ferrara}  \& {White}}{{Ciardi}
  et~al.}{2003}]{Ciardi2003}
{Ciardi} B.,  {Ferrara} A.,   {White} S.~D.~M.,  2003, \mn@doi [\mnras]
  {10.1046/j.1365-8711.2003.06976.x}, \href
  {http://adsabs.harvard.edu/abs/2003MNRAS.344L...7C} {344, L7}

\bibitem[\protect\citeauthoryear{Datta, Mellema, Mao, Iliev, Shapiro  \&
  Ahn}{Datta et~al.}{2012}]{Datta2012}
Datta K.~K.,  Mellema G.,  Mao Y.,  Iliev I.~T.,  Shapiro P.~R.,   Ahn K.,
  2012, \mn@doi [MNRAS] {10.1111/j.1365-2966.2012.21293.x}, 424, 1877

\bibitem[\protect\citeauthoryear{Datta, Jensen, Majumdar, Mellema, Iliev, Mao,
  Shapiro  \& Ahn}{Datta et~al.}{2014}]{Datta2014}
Datta K.~K.,  Jensen H.,  Majumdar S.,  Mellema G.,  Iliev I.~T.,  Mao Y.,
  Shapiro P.~R.,   Ahn K.,  2014, \mn@doi [MNRAS] {10.1093/mnras/stu927}, 442,
  1491

\bibitem[\protect\citeauthoryear{{Fan} et~al.,}{{Fan} et~al.}{2006}]{fan06b}
{Fan} X.,  et~al., 2006, \mn@doi [\aj] {10.1086/504836}, \href
  {https://ui.adsabs.harvard.edu/abs/2006AJ....132..117F} {132, 117}

\bibitem[\protect\citeauthoryear{{Finlator}, {{\"O}zel}  \&
  {Dav{\'e}}}{{Finlator} et~al.}{2009}]{Finlator2009}
{Finlator} K.,  {{\"O}zel} F.,   {Dav{\'e}} R.,  2009, \mn@doi [\mnras]
  {10.1111/j.1365-2966.2008.14190.x}, \href
  {http://adsabs.harvard.edu/abs/2009MNRAS.393.1090F} {393, 1090}

\bibitem[\protect\citeauthoryear{{Furlanetto}, {Oh}  \& {Briggs}}{{Furlanetto}
  et~al.}{2006}]{furlanetto06a}
{Furlanetto} S.~R.,  {Oh} S.~P.,   {Briggs} F.~H.,  2006, \mn@doi [\physrep]
  {10.1016/j.physrep.2006.08.002}, \href
  {https://ui.adsabs.harvard.edu/abs/2006PhR...433..181F} {433, 181}

\bibitem[\protect\citeauthoryear{Ghara, Choudhury  \& Datta}{Ghara
  et~al.}{2015}]{Ghara2015}
Ghara R.,  Choudhury T.~R.,   Datta K.~K.,  2015, \mn@doi [MNRAS]
  {10.1093/mnras/stu2512}, 447, 1806

\bibitem[\protect\citeauthoryear{{Gnedin}}{{Gnedin}}{2000}]{Gnedin2000}
{Gnedin} N.~Y.,  2000, \mn@doi [\apj] {10.1086/317042}, \href
  {http://adsabs.harvard.edu/abs/2000ApJ...542..535G} {542, 535}

\bibitem[\protect\citeauthoryear{{Iliev}, {Mellema}, {Ahn}, {Shapiro}, {Mao}
  \& {Pen}}{{Iliev} et~al.}{2014}]{Iliev2014}
{Iliev} I.~T.,  {Mellema} G.,  {Ahn} K.,  {Shapiro} P.~R.,  {Mao} Y.,   {Pen}
  U.-L.,  2014, \mn@doi [\mnras] {10.1093/mnras/stt2497}, \href
  {http://adsabs.harvard.edu/abs/2014MNRAS.439..725I} {439, 725}

\bibitem[\protect\citeauthoryear{Jensen, Datta, Mellema, Chapman, Abdalla  \&
  {and the LOFAR-EoR Group,}}{Jensen et~al.}{2013}]{jensen13}
Jensen H.,  Datta K.~K.,  Mellema G.,  Chapman E.,  Abdalla F.~B.,   {and the
  LOFAR-EoR Group,} 2013, \mn@doi [MNRAS] {10.1093/mnras/stt1341}, 435, 460

\bibitem[\protect\citeauthoryear{{La Plante}, {Battaglia}, {Natarajan},
  {Peterson}, {Trac}, {Cen}  \& {Loeb}}{{La Plante} et~al.}{2014}]{laplante14}
{La Plante} P.,  {Battaglia} N.,  {Natarajan} A.,  {Peterson} J.~B.,  {Trac}
  H.,  {Cen} R.,   {Loeb} A.,  2014, \mn@doi [\apj]
  {10.1088/0004-637X/789/1/31}, \href
  {https://ui.adsabs.harvard.edu/abs/2014ApJ...789...31L} {789, 31}

\bibitem[\protect\citeauthoryear{Majumdar, Bharadwaj  \& Choudhury}{Majumdar
  et~al.}{2012}]{majumdar12}
Majumdar S.,  Bharadwaj S.,   Choudhury T.~R.,  2012, \mn@doi [MNRAS]
  {10.1111/j.1365-2966.2012.21914.x}, 426, 3178

\bibitem[\protect\citeauthoryear{Majumdar, Bharadwaj  \& Choudhury}{Majumdar
  et~al.}{2013}]{Majumdar2013}
Majumdar S.,  Bharadwaj S.,   Choudhury T.~R.,  2013, \mn@doi [MNRAS]
  {10.1093/mnras/stt1144}, 434, 1978

\bibitem[\protect\citeauthoryear{Majumdar et~al.,}{Majumdar
  et~al.}{2016}]{Majumdar2015}
Majumdar S.,  et~al., 2016, \mn@doi [MNRAS] {10.1093/mnras/stv2812}, 456, 2080

\bibitem[\protect\citeauthoryear{Mao, Shapiro, Mellema, Iliev, Koda  \&
  Ahn}{Mao et~al.}{2012}]{mao12}
Mao Y.,  Shapiro P.~R.,  Mellema G.,  Iliev I.~T.,  Koda J.,   Ahn K.,  2012,
  \mn@doi [MNRAS] {10.1111/j.1365-2966.2012.20471.x}, 422, 926

\bibitem[\protect\citeauthoryear{{Mellema}, {Iliev}, {Alvarez}  \&
  {Shapiro}}{{Mellema} et~al.}{2006a}]{Mellema2006}
{Mellema} G.,  {Iliev} I.~T.,  {Alvarez} M.~A.,   {Shapiro} P.~R.,  2006a,
  \mn@doi [\na] {10.1016/j.newast.2005.09.004}, \href
  {http://adsabs.harvard.edu/abs/2006NewA...11..374M} {11, 374}

\bibitem[\protect\citeauthoryear{Mellema, Iliev, Pen  \& Shapiro}{Mellema
  et~al.}{2006b}]{mellema06b}
Mellema G.,  Iliev I.~T.,  Pen U.-L.,   Shapiro P.~R.,  2006b, \mn@doi [MNRAS]
  {10.1111/j.1365-2966.2006.10919.x}, 372, 679

\bibitem[\protect\citeauthoryear{{Mellema} et~al.,}{{Mellema}
  et~al.}{2013}]{2013ExA....36..235M}
{Mellema} G.,  et~al., 2013, \mn@doi [Experimental Astronomy]
  {10.1007/s10686-013-9334-5}, \href
  {http://adsabs.harvard.edu/abs/2013ExA....36..235M} {36, 235}

\bibitem[\protect\citeauthoryear{{Mesinger} \& {Furlanetto}}{{Mesinger} \&
  {Furlanetto}}{2007}]{Mesinger2007}
{Mesinger} A.,  {Furlanetto} S.,  2007, \mn@doi [\apj] {10.1086/521806}, \href
  {http://adsabs.harvard.edu/abs/2007ApJ...669..663M} {669, 663}

\bibitem[\protect\citeauthoryear{Mesinger, Furlanetto  \& Cen}{Mesinger
  et~al.}{2011}]{mesinger11}
Mesinger A.,  Furlanetto S.,   Cen R.,  2011, \mn@doi [MNRAS]
  {10.1111/j.1365-2966.2010.17731.x}, 411, 955

\bibitem[\protect\citeauthoryear{Mihalas}{Mihalas}{1978}]{mihalas78}
Mihalas D.,  1978, Stellar Atmospheres.
A Series of books in astronomy and astrophysics, W. H. Freeman

\bibitem[\protect\citeauthoryear{{Mondal}, {Bharadwaj}  \& {Datta}}{{Mondal}
  et~al.}{2018}]{Mondal2017}
{Mondal} R.,  {Bharadwaj} S.,   {Datta} K.~K.,  2018, \mn@doi [\mnras]
  {10.1093/mnras/stx2888}, \href
  {https://ui.adsabs.harvard.edu/abs/2018MNRAS.474.1390M} {474, 1390}

\bibitem[\protect\citeauthoryear{{Mortlock} et~al.,}{{Mortlock}
  et~al.}{2011}]{Mortlock2011}
{Mortlock} D.~J.,  et~al., 2011, \mn@doi [\nat] {10.1038/nature10159}, \href
  {http://adsabs.harvard.edu/abs/2011Natur.474..616M} {474, 616}

\bibitem[\protect\citeauthoryear{{Planck Collaboration} et~al.,}{{Planck
  Collaboration} et~al.}{2018}]{Aghanim2018}
{Planck Collaboration} et~al., 2018, arXiv e-prints, \href
  {https://ui.adsabs.harvard.edu/abs/2018arXiv180706209P} {p. arXiv:1807.06209}

\bibitem[\protect\citeauthoryear{{Pritchard} \& {Loeb}}{{Pritchard} \&
  {Loeb}}{2010}]{pritchard10a}
{Pritchard} J.~R.,  {Loeb} A.,  2010, \mn@doi [\prd]
  {10.1103/PhysRevD.82.023006}, \href
  {https://ui.adsabs.harvard.edu/abs/2010PhRvD..82b3006P} {82, 023006}

\bibitem[\protect\citeauthoryear{{Santos} \& {Cooray}}{{Santos} \&
  {Cooray}}{2006}]{santos06}
{Santos} M.~G.,  {Cooray} A.,  2006, \mn@doi [\prd]
  {10.1103/PhysRevD.74.083517}, \href
  {http://adsabs.harvard.edu/abs/2006PhRvD..74h3517S} {74, 083517}

\bibitem[\protect\citeauthoryear{{Santos}, {Amblard}, {Pritchard}, {Trac},
  {Cen}  \& {Cooray}}{{Santos} et~al.}{2008}]{2008ApJ...689....1S}
{Santos} M.~G.,  {Amblard} A.,  {Pritchard} J.,  {Trac} H.,  {Cen} R.,
  {Cooray} A.,  2008, \mn@doi [\apj] {10.1086/592487}, \href
  {http://adsabs.harvard.edu/abs/2008ApJ...689....1S} {689, 1}

\bibitem[\protect\citeauthoryear{{Santos}, {Ferramacho}, {Silva}, {Amblard}  \&
  {Cooray}}{{Santos} et~al.}{2010}]{santos2010}
{Santos} M.~G.,  {Ferramacho} L.,  {Silva} M.~B.,  {Amblard} A.,   {Cooray} A.,
   2010, \mn@doi [\mnras] {10.1111/j.1365-2966.2010.16898.x}, \href
  {http://adsabs.harvard.edu/abs/2010MNRAS.406.2421S} {406, 2421}

\bibitem[\protect\citeauthoryear{{Santos}, {Silva}, {Pritchard}, {Cen}  \&
  {Cooray}}{{Santos} et~al.}{2011}]{2011A&A...527A..93S}
{Santos} M.~G.,  {Silva} M.~B.,  {Pritchard} J.~R.,  {Cen} R.,   {Cooray} A.,
  2011, \mn@doi [\aap] {10.1051/0004-6361/201015695}, \href
  {http://adsabs.harvard.edu/abs/2011A%26A...527A..93S} {527, A93}

\bibitem[\protect\citeauthoryear{{Semelin}, {Combes}  \& {Baek}}{{Semelin}
  et~al.}{2007}]{Semelin2007}
{Semelin} B.,  {Combes} F.,   {Baek} S.,  2007, \mn@doi [\aap]
  {10.1051/0004-6361:20077965}, \href
  {http://adsabs.harvard.edu/abs/2007A%26A...474..365S} {474, 365}

\bibitem[\protect\citeauthoryear{Shapiro, Mao, Iliev, Mellema, Datta, Ahn  \&
  Koda}{Shapiro et~al.}{2013}]{shapiro13}
Shapiro P.~R.,  Mao Y.,  Iliev I.~T.,  Mellema G.,  Datta K.~K.,  Ahn K.,
  Koda J.,  2013, \mn@doi [Phys. Rev. Lett.] {10.1103/PhysRevLett.110.151301},
  110, 151301

\bibitem[\protect\citeauthoryear{{Trac} \& {Cen}}{{Trac} \&
  {Cen}}{2007}]{Trac2007}
{Trac} H.,  {Cen} R.,  2007, \mn@doi [\apj] {10.1086/522566}, \href
  {http://adsabs.harvard.edu/abs/2007ApJ...671....1T} {671, 1}

\bibitem[\protect\citeauthoryear{Trott}{Trott}{2016}]{trott_lc_16}
Trott C.~M.,  2016, \mn@doi [MNRAS] {10.1093/mnras/stw1310}, 461, 126

\bibitem[\protect\citeauthoryear{{Zahn}, {Lidz}, {McQuinn}, {Dutta},
  {Hernquist}, {Zaldarriaga}  \& {Furlanetto}}{{Zahn} et~al.}{2007}]{Zahn2007}
{Zahn} O.,  {Lidz} A.,  {McQuinn} M.,  {Dutta} S.,  {Hernquist} L.,
  {Zaldarriaga} M.,   {Furlanetto} S.~R.,  2007, \mn@doi [\apj]
  {10.1086/509597}, \href {http://adsabs.harvard.edu/abs/2007ApJ...654...12Z}
  {654, 12}

\bibitem[\protect\citeauthoryear{Zawada, Semelin, Vonlanthen, Baek  \&
  Revaz}{Zawada et~al.}{2014}]{Zawada2014}
Zawada K.,  Semelin B.,  Vonlanthen P.,  Baek S.,   Revaz Y.,  2014, \mn@doi
  [MNRAS] {10.1093/mnras/stu035}, 439, 1615

\makeatother
\end{thebibliography}
\end{document}